\documentclass[paper]{JHEP3}

\usepackage{amsmath}
\usepackage{amsfonts}
\usepackage{amssymb}
\usepackage{graphicx}

\newcommand{\mfig}{Figure }
\newcommand{\msec}{section }

\newcommand{\nvec}[1]{\boldsymbol{#1}}
\newcommand{\abs}[1]{\left\vert #1 \right\vert}
\newcommand{\ud}{\mathrm{d}}

\newcommand{\rt}{r}
\newcommand{\vrt}{\nvec{r}}

\newcommand{\sigmadip}{\hat{\sigma}}
\newcommand{\sigmatot}{\sigma}

\preprint{ZU-TH 21/10}
\title{On the Energy Dependence of the Dipole-Proton Cross Section in 
Deep Inelastic Scattering}
\author{Carlo Ewerz\,$^{a,b,1}$, Andreas von Manteuffel\,$^{c,2}$, Otto Nachtmann\,$^{a,3}$
\\
$^a$
Institut f\"ur Theoretische Physik, Universit\"at Heidelberg,\\
\phantom{$^a$}
Philosophenweg 16, D-69120 Heidelberg, Germany\\
$^b$
ExtreMe Matter Institute EMMI, GSI Helmholtzzentrum f\"ur Schwerionenforschung,\\
\phantom{$^a$} 
Planckstra{\ss}e 1, D-64291 Darmstadt, Germany\\
$^c$
Institut f\"ur Theoretische Physik, Universit\"at Z\"urich, \\
\phantom{$^a$}
Winterthurerstr.\ 190, CH-8057 Z\"urich, Switzerland
\\
$^1$E-mail: \email{C.Ewerz@thphys.uni-heidelberg.de}\\
$^2$E-mail: \email{manteuffel@physik.uzh.ch}\\
$^3$E-mail: \email{O.Nachtmann@thphys.uni-heidelberg.de}
}

\abstract{
We study the dipole picture of high-energy virtual-photon-proton 
scattering. It is shown that different choices for the energy 
variable in the dipole cross section used in the literature 
are not related to each other by simple 
arguments equating the typical dipole size and the inverse photon 
virtuality, contrary to what is often stated. We argue that the good 
quality of fits to structure functions that use 
Bjorken-$x$ as the energy variable -- which is strictly speaking 
not justified in the dipole picture -- can instead be understood as a consequence 
of the sign of scaling violations that occur for increasing $Q^2$ at fixed small $x$. 
We show that the dipole formula for massless quarks has the structure 
of a convolution. From this we obtain derivative relations between the structure 
function $F_2$ at large and small $Q^2$ and the dipole-proton cross section 
at small and large dipole size $r$, respectively. 
}


\begin{document}

\section{Introduction}
\label{sec:intro}

The colour dipole model \cite{Nikolaev:1990ja,Nikolaev:et,Mueller:1993rr}
provides a successful description of deep inelastic
scattering (DIS) processes in a wide range of the kinematic variables. 
We consider here electron- and positron-proton scattering, 
\begin{equation} 
\label{eptoeX}
e^\pm + p \to e^\pm + X \,,
\end{equation}
at not too high momentum transfers squared 
$Q^2$, $Q^2 \lesssim 1000~\mbox{GeV}^2$ say, and 
at high energies. There, only the exchange of a virtual photon $\gamma^*$ 
between the leptons and the hadrons has to be taken into account. 
Thus, we study in essence the absorption of a high-energy virtual photon 
on a proton, 
\begin{equation}
\label{processapintro}
\gamma^\ast + p \rightarrow X
\,.
\end{equation}
The structure functions for this inclusive DIS process were extensively 
measured at HERA 
\cite{Breitweg:2000yn,Adloff:2000qk,Chekanov:2001qu,Adloff:2003uh,Chekanov:2003yv,aaron:2009wt}. 
A considerable part of the literature concerning the proton structure functions 
in DIS uses the colour dipole picture as an essential input. In particular, the 
interpretation of the structure function $F_2$ at small Bjorken-$x$ relies 
heavily on the dipole picture. The corresponding results are frequently used 
in the context of various scattering processes in proton-proton collisions 
at the LHC and in the description of the initial conditions of the creation of 
the quark-gluon plasma in heavy-ion collisions. Also calculations for processes 
at a future lepton-hadron collider \cite{eicwhitepaper} are often based on the dipole picture. 
Obviously, a good understanding of the dipole picture and its consequences is 
very important for all these applications. In the present study we shall consider 
an aspect of the dipole picture that has -- in our opinion -- not yet received 
proper attention in the literature, namely the correct choice of energy 
variable in the dipole-proton cross section.  

The idea that the high energy photon in the reaction \eqref{processapintro} 
acts in some way like a hadron goes back a long time.
It has been used for instance since the 1960s in vector dominance 
models, see \cite{Bauer:1977iq,Schildknecht:2005xr} for reviews. 
Today, the dipole picture is frequently used in analyses of DIS structure functions. 
There, the reaction \eqref{processapintro} is viewed as a two-step process.
In the first step the photon splits into a quark-antiquark pair which represents 
the colour dipole. Subsequently, that pair scatters on the proton, this second 
step being a purely hadronic reaction. 
For reviews we refer the reader to \cite{Donnachie:2002en,Close:2007zzd}.
In \cite{Ewerz:2004vf,Ewerz:2006vd} the foundations of this dipole picture 
were examined in detail. The precise assumptions which have to be made in 
order to arrive at it were spelled out. In \cite{Ewerz:2006an,Ewerz:2007md} it 
was shown that already the general formulae of the standard dipole approach 
allow one to derive stringent bounds on various ratios of structure functions.
These bounds were used to determine the kinematic region where the dipole
picture is possibly applicable.
In particular, it was found that for $\gamma^\ast p$ c.\,m.\ energies $W$ in the
range 60 to 240~GeV the standard dipole picture fails to be compatible with the
HERA data for $Q^2$ larger than about $100$ to $200$~GeV$^2$, 
see Fig.~9 of \cite{Ewerz:2007md}.

The derivations of some of these bounds rely on the dipole-proton 
cross sections $\hat{\sigma}^{(q)}$, where $q$ denotes the quark flavour, 
being independent of $Q^2$. In \cite{Ewerz:2006vd} it has been stressed that 
this $Q^2$-independence of the dipole-proton cross section is in fact natural. 
The correct energy variable is given exclusively by $W$ and the functional 
dependence of $\hat{\sigma}^{(q)}$ should be 
\begin{equation}
\label{correctvardep}
\sigmadip^{(q)} = \sigmadip^{(q)}(r,W)\,,
\end{equation}
where $r$ is the transverse size of the dipole. 
This excludes in particular the choice of Bjorken-$x$ instead of $W$, since
this would introduce a dependence on $Q^2$ in addition to $W$, see \eqref{defkin} below.
In the derivation \cite{Ewerz:2004vf,Ewerz:2006vd} of the dipole picture
the dipole cross section arises
from a $T$-matrix element for the scattering of a dipole state on the proton.
The key feature of these dipole states is that they consist
of a quark and an antiquark described by asymptotic states.
The dipole states are then independent of $Q^2$ in the high energy limit.
Upon a smearing in the relative transverse vector $\vrt$ between quark and 
antiquark and in the longitudinal momentum fraction $\alpha$ of the 
photon carried by the quark the dipole states can be viewed as hadron analogues,
whose normalisation is independent of continuous internal degrees of freedom.
But also the mean squared invariant mass of such smeared dipole states
is independent of $Q^2$ at large $W$.
Since any physical cross section can depend only on variables defined 
by the incoming states, the dipole-proton cross section hence cannot 
depend on $Q^2$. 
Nevertheless, the energy variable $x$ -- and therefore a $Q^2$-dependence --
is frequently used in popular models for the dipole cross section, such as 
\cite{GolecBiernat:1998js}. Further examples for $x$-dependent dipole 
cross sections are \cite{Bartels:2002cj}, \cite{Iancu:2003ge}, and \cite{Albacete:2009fh}. 
Sometimes also other dependencies on $Q^2$ are introduced by modifying 
the photon wave functions \cite{Dosch:1997nw,Forshaw:1999uf}. 
Other models introduce an impact parameter dependence of the dipole cross section, 
see \cite{Kowalski:2003hm,Watt:2007nr}. 
For an overview of the physical motivations for various models see 
for instance \cite{Motyka:2008jk}. 
Furthermore, the perturbative gluon density $g(x,Q^2)$ naturally depends on
$x$ and $Q^2$ and its often assumed proportionality to the
dipole cross section suggests that the latter is also $Q^2$-dependent.
A subtle point in such a comparison is given by the fact that
different limits are used for the dipole picture and the
double leading logarithmic approximation, respectively, see the discussion in
\cite{Ewerz:2006vd}.
Only very few models for the dipole cross section have been constructed that 
use the correct functional dependence \eqref{correctvardep}, among them are 
\cite{Forshaw:1999uf}, \cite{Cvetic:2001ie}, and \cite{Donnachie:2001wt}. 

Through the interplay of photon wave function and dipole cross section
a typical transverse dipole size is generated, which depends on $Q$.
In this paper we shall investigate whether such an effective dipole size 
can be used to relate different choices of energy variables in the dipole cross section.

We consider then the dipole formulae for the case of massless quarks and 
show that these formulae can be understood as a convolution. This is used 
to analyse the relation of the structure functions and the dipole cross section. 
For high and for low $Q^2$ we find simple but somewhat surprising relations. 

Our paper is organised as follows.
Section \ref{sec:picture} reviews the relevant formulae of the dipole picture. 
In section \ref{sec:dipenergydep} we discuss typical dipole sizes 
and investigate whether different choices for the energy dependence of 
the dipole cross section may be related to each other by effective scale 
arguments. In section \ref{sec:convolution} we rewrite the dipole formula 
for massless quarks as a convolution. We present considerations 
suggesting that the success of using Bjorken-$x$ as the energy variable 
in the dipole cross section is due to the specific form of scaling violations 
in the structure function $F_2$. We derive asymptotic relations for the 
general dipole picture in the regimes of large and small $Q^2$, 
respectively. Our conclusions are drawn in section \ref{sec:conclusions}. 
Two appendices contain results used in 
the main text: In appendix \ref{appA} we derive the asymptotic behaviour 
of the integrated photon densities. Appendix \ref{appB} 
explains the steps for obtaining a simplified version of a $W$-dependent 
model for the dipole-proton cross section from the original model 
\cite{Donnachie:2001wt}. 

\section{The dipole picture}
\label{sec:picture}

We use the standard formulae for the kinematics and for the definitions of
structure functions of the reaction \eqref{eptoeX}, 
see for instance \cite{Nachtmann:1990ta}.
As discussed above, we consider $Q^2 \lesssim 1000~\text{GeV}^2$ 
so that it is sufficient to take into account the exchange of a photon. 
Thus, we shall study in the following the absorption of a 
virtual photon $\gamma^\ast$ on the proton,
\begin{equation}\label{processap}
\gamma^\ast(q) + p(p) \rightarrow X(p')\,.
\end{equation}
Here the $4$-momenta are indicated in brackets.
The c.\,m.\ energy for this reaction is denoted by $W$, the virtuality of
$\gamma^\ast$ by $Q^2$.
For these and the other usual variables we have 
\begin{align}
\label{defkin}
W^2 &= (p+q)^2\,, \notag\\
Q^2 &= -q^2\,,\notag\\
\nu &= \frac{p \cdot q}{2 m_p}\,,\notag\\
x &= \frac{Q^2}{2 m_p \nu} = \frac{Q^2}{W^2+Q^2-m_p^2}\,.
\end{align}
The proton in \eqref{processap} is supposed to be unpolarised, while the virtual photon
can have transverse or longitudinal polarisation.
The corresponding total cross sections are $\sigma_T(W,Q^2)$ and
$\sigma_L(W,Q^2)$, respectively.
The $F_2$ structure function is, with Hand's convention \cite{Hand:1963bb} 
for the $\gamma^\ast$ flux factor, 
\begin{equation}
\label{f2diphand}
F_2(W,Q^2) = \frac{Q^2}{4 \pi^2 \alpha_{\rm em}}
   \left[ \sigma_T(W,Q^2) + \sigma_L(W,Q^2) \right] (1-x)
+ \mathcal{O}\left(\frac{m_p^2}{W^2}\right) \,.
\end{equation}
For small Bjorken-$x$, $x\ll 1$, this simplifies to
\begin{equation}\label{f2dipsimple}
F_2(W,Q^2) = \frac{Q^2}{4\pi^2\alpha_{\mathrm{em}}}
  \left[ \sigma_T(W,Q^2) + \sigma_L(W,Q^2) \right]\,.
\end{equation}
In the following we shall use this simpler relation since we shall only consider
the region $x\ll 1$.

In order to obtain the standard dipole model for the cross sections $\sigma_{T,L}$ 
we can relate them first to the imaginary part of the
$\gamma^\ast p \to \gamma^\ast p$ forward scattering amplitude.
The latter is represented as the initial $\gamma^\ast$ splitting into a $q\bar q$
pair, this pair scattering on the proton and the $q\bar q$ subsequently fusing
into the final state $\gamma^\ast$, see Figure \ref{fig:dipolediag}.%
\FIGURE[ht]{
\includegraphics[width=0.7\textwidth]{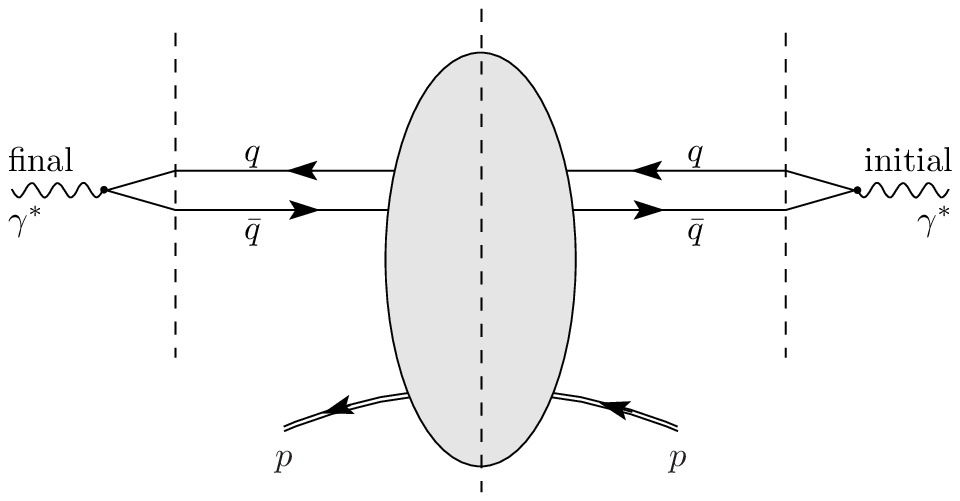}
\caption{\label{fig:dipolediag}
Basic diagram for the description of the cross sections $\sigma_{T,L}$ of
$\gamma^\ast p$ scattering in the standard dipole approach. 
}}
With the assumptions spelled out in detail in \msec 6 of \cite{Ewerz:2006vd} 
the diagram of \mfig \ref{fig:dipolediag} gives in the high energy limit 
\begin{equation}
\label{sigmatdip}
\sigmatot_{T,L}(W,Q^2)=
\sum_q\int \ud^2 \rt \,
w^{(q)}_{T,L}(\rt,Q^2)\,
\sigmadip^{(q)}(\rt,W) \,.
\end{equation}
Here the integrated `photon densities' are 
\begin{align}
\label{denst}
w^{(q)}_T(\rt,Q^2) &=
\sum_{\lambda,\lambda'}\int^1_0 \ud \alpha \,
\left|
\psi^{(q)+}_{\lambda\lambda'}(\alpha,\rt,Q) \right|^2 
\,,
\\
\label{densl}
w^{(q)}_L(\rt,Q^2) &=
\sum_{\lambda,\lambda'}\int^1_0 \ud \alpha\,
\left| 
\psi^{(q)L}_{\lambda\lambda'}(\alpha,\rt,Q)\right|^2 \,. 
\end{align}
We recall that $\alpha$ is the longitudinal momentum fraction of the 
$\gamma^\ast$
carried by the quark, $\vrt$ is the vector in transverse position space
from the antiquark to the quark, $r=|\vrt|$, 
and $\lambda$ and $\lambda'$ are the 
helicities of $q$ and $\bar q$, respectively. 
The total cross section for the scattering of the $q\bar q$ pair on the proton
is denoted by $\sigmadip^{(q)}$,
the $\gamma^\ast$ wave functions for transversely and longitudinally polarised
$\gamma^\ast$ by $\psi_{\lambda\lambda'}^{(q)\,\pm}$ and
$\psi_{\lambda\lambda'}^{(q)\,L}$, respectively.
A sum over all contributing quark flavours $q$ is to be performed in \eqref{sigmatdip}. 
Inserting in \eqref{denst} and \eqref{densl} the photon wave functions 
in leading order of $\alpha_s$ and assuming the longitudinal momenta 
of quark and antiquark to be much larger than their mass and transverse 
momenta we get the standard expressions 
\begin{align}
\label{sumpsitdens}
        \sum_{\lambda, \lambda'} \left| 
        \psi_{\lambda \lambda'}^{(q) +} 
          (\alpha, \vrt,Q) \right|^2 
&=
        \frac{N_c}{2 \pi^2} \, \alpha_{\rm em} Q_q^2 
        \left\{  \left[ \alpha^2 + (1-\alpha)^2 \right] 
        \epsilon_q^2 [K_1(\epsilon_q \rt) ]^2 
+ m_q^2 [K_0(\epsilon_q \rt) ]^2 
        \right\}  \,,
\\
\label{sumpsildens}
        \sum_{\lambda, \lambda'} \left| 
        \psi_{\lambda \lambda'}^{(q) L}(\alpha, \vrt,Q) 
        \right|^2 
&=
        \frac{2 N_c}{\pi^2} \, \alpha_{\rm em} Q_q^2
         Q^2 [\alpha (1-\alpha)]^2 [K_0(\epsilon_q \rt) ]^2  \,.
\end{align}
Here $N_c=3$ is the number of colours, $Q_q$ is the charge of the 
quark in units of the proton charge, $K_{0,1}$ are modified Bessel 
functions, and 
\begin{equation}
\label{defepsq}
\epsilon_q = \sqrt{\alpha (1-\alpha) Q^2 + m_q^2} \,.
\end{equation}
For massless quarks $q$ the integrated photon densities 
$w^{(q)}_{T,L}$ of \eqref{denst} and \eqref{densl} can be calculated analytically. 
The corresponding formulae are given in appendix \ref{appA}. 

In \eqref{sigmatdip} we have written $\sigma_T$ and $\sigma_L$ as 
functions of $W$ and $Q^2$, and the dipole cross sections 
$\sigmadip^{(q)}$ as functions of $r$ and $W$. It is one purpose 
of this paper to examine the consequences of choosing energy 
variables other than $W$ in the dipole formulae. Choosing, 
for instance, Bjorken-$x$ as variable we get again the 
formulae \eqref{sigmatdip} but with $W$ everywhere replaced 
by $x$. On the l.h.s.\ of \eqref{sigmatdip} this means, of course, 
just another and equivalent pair of variables. But for the r.h.s.\ 
this has drastic consequences as we shall show in the following. 

\section{Energy dependence of the dipole cross section}
\label{sec:dipenergydep}

\subsection{Typical dipole sizes}
\label{sec:diptypical}

In \cite{Ewerz:2006vd} it was shown that in a formulation starting from 
the functional integral the dipole cross section $\sigmadip^{(q)}$ comes 
out as independent of $Q^2$ and depends on $r$ and the energy $W$ only.
We already noted that in contrast to this, prominent dipole models discussed in
the literature introduce a dependence of $\sigmadip^{(q)}$ on $Q^2$, typically
through the Bjorken-$x$ variable.
Often it is argued that in the dipole model one has a relation of the kind
\begin{equation}
\label{naivedipsize}
\rt =\frac{C}{Q}\,,
\end{equation}
with a constant $C$, corresponding to a `typical dipole size' or `typical scale'. 
That is, the most relevant dipole sizes $\rt$ are determined by the scale $Q$.
The reasoning behind this is the fact that due to the interplay of the 
$Q^2$-dependence of the photon wave function and the $\rt$-dependence of the
dipole cross section a typical size is generated.
Neglecting the $W$-dependence, the masses and all non-perturbative intrinsic
scales, the typical size must be given by \eqref{naivedipsize} for dimensional 
reasons. 
Taking \eqref{naivedipsize} seriously one might be tempted to replace 
freely in $\sigmadip (\rt,W)$ or related quantities $\rt$ by $Q^2$ 
dependencies and vice versa. In this section we shall show that such 
replacements are far from harmless and, in fact, are not admissible. 

Let us first see how the relation \eqref{naivedipsize} for typical
dipole sizes arises. Inserting \eqref{sigmatdip} into \eqref{f2dipsimple} 
we can write the dipole model expression for the structure function $F_2$ 
as 
\begin{equation}
\label{f2withh}
F_2(W,Q^2) = Q \int_0^\infty \ud r \sum_q h(Q r, m_q r) \, \frac{1}{r^2}\, Q_q^2 
\sigmadip^{(q)} (r,W) \,.
\end{equation}
Here we define 
\begin{equation}
\label{defh}
h(Q r, m_q r) = \frac{Q r^3}{2 \pi \alpha_\mathrm{em} Q_q^2}\,
\left[ w_T^{(q)} (r,Q^2) + w_L^{(q)} (r,Q^2) \right] \,.
\end{equation}
The dependence of the dimensionless functions $Q r^3 w^{(q)}_{T,L}/(\alpha_{\rm em} Q_q^2)$ 
on $r$, as resulting from \eqref{denst}-\eqref{sumpsildens}, is shown 
in Figure \ref{fig:spikes} for the case of massless quarks.%
\FIGURE[ht]{
\hspace*{-1cm}
\includegraphics[width=1.05 \textwidth]{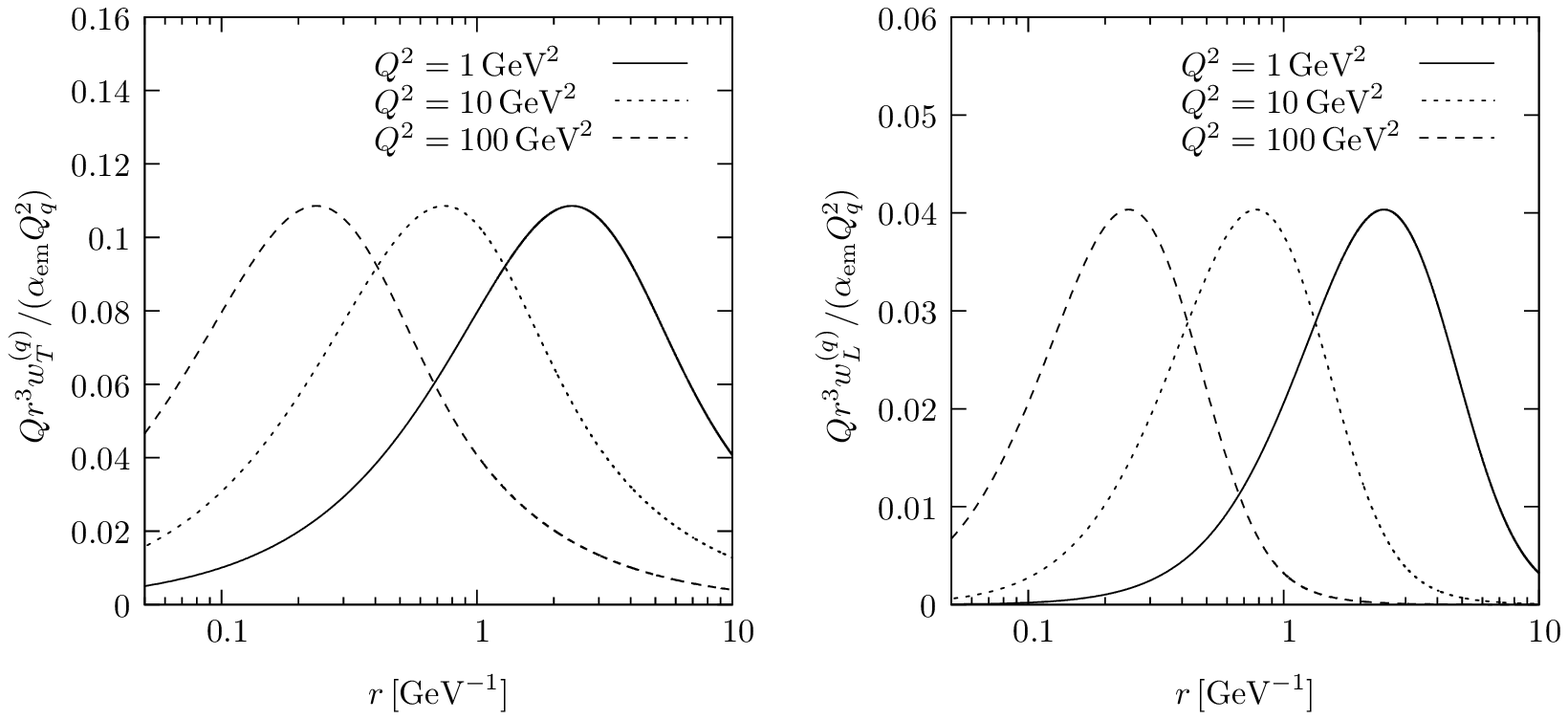}
\caption{
Dependence of the dimensionless quantity 
$Q r^3 w^{(q)}_{T,L}/(\alpha_{\rm em} Q_q^2)$ on $r$ for massless ($q=u,d,s$) quarks 
for three different values $Q^2=1,10,100\,\mbox{GeV}^2$; 
on the left for transversely polarised photons, on the right for 
longitudinally polarised photons. 
\label{fig:spikes}
}}%
The function $h$ is shown in Figure \ref{fig:Fighz0}. 
For massless quarks, $m_q=0$, $h$ is only a function of the 
dimensionless variable 
\begin{equation}
\label{zQr}
z = Q r \,.
\end{equation}
The function $h(z,0)$ has its maximum at 
\begin{equation}
\label{z0}
z_0 = 2.4010 \,.
\end{equation}
\FIGURE[ht]{
\includegraphics[width=0.6\textwidth]{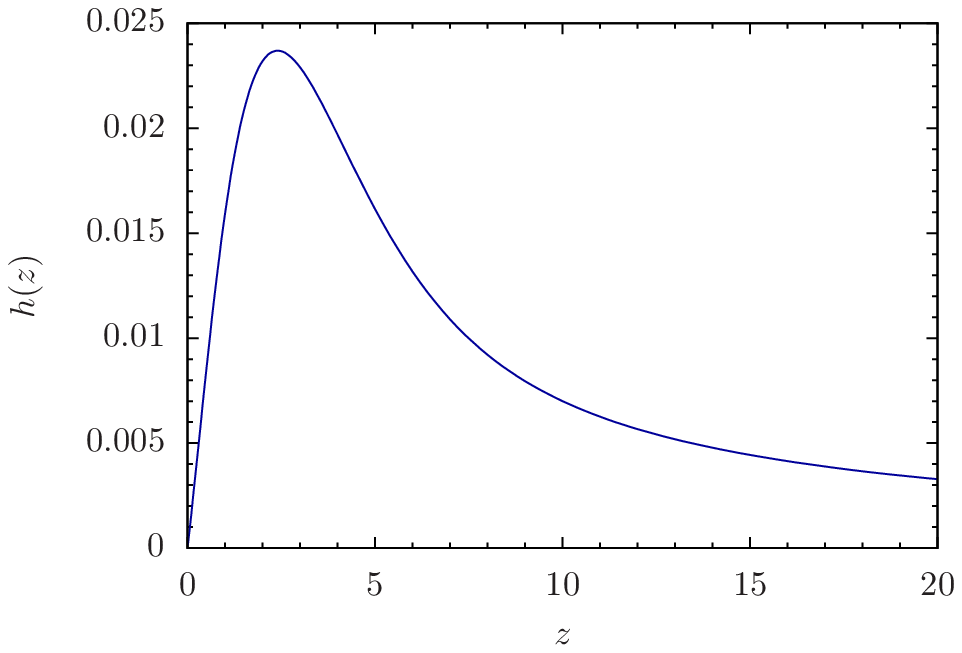}
\caption{\label{fig:Fighz0}
The function $h(z,0)$ defined in \eqref{defh}. 
}}
The behaviour of $h(z,0)$ for small and large $z$ is 
\begin{equation}
\label{hbehav}
\begin{split}
&h(z,0) \propto z \quad \quad \quad \:\mbox{for} \:\:z \to 0 \,, \\
&h(z,0) \propto \frac{1}{z} \quad \quad \quad \mbox{for} \:\:z \to \infty \,. 
\end{split}
\end{equation}
In many proposed models colour transparency at small $r$ is implemented 
by assuming $\sigmadip^{(q)} (r) \propto r^2$ for $r \to 0$. For larger $r$, 
on the other hand, the dipole cross sections are certainly not expected to 
grow faster than $r^2$. Thus $Q_q^2 \sigmadip^{(q)} (r) /r^2$ should be a rather smooth function of $r$. 
Let us now, for the sake of argument, consider massless quarks only. 
Then we get from \eqref{f2withh} 
\begin{equation}
\label{Fwithhf}
F_2(W,Q^2) = Q \int_0^\infty \ud r \,h(Qr,0) \,f\left(r,\frac{1}{r^2W^2}\right)
\,,
\end{equation}
where we define the function $f$ by 
\begin{equation}
\label{deff}
\sum_q Q_q^2 \sigmadip^{(q)} (r,W) = r^2 f\left(r,\frac{1}{r^2W^2}\right) \,.
\end{equation}
In \eqref{Fwithhf} a smooth function of $r$, $f(r,1/(r^2W^2))$, 
is integrated with $h(Qr,0)$ having a maximum at $r_\textrm{max}=z_0/Q$. 
It is now tempting to replace $r$ in the smooth function $f(r,1/(r^2W^2))$ 
by $r_\textrm{max}$. In this way we get a modified $F_2$, 
\begin{equation}
\label{F2modi}
F_2^{\mathrm{mod}'}(W,Q^2)= 
Q \int_0^\infty \ud r \,h(Qr,0) \,f\left(\frac{z_0}{Q}, \frac{x}{z_0^2} \right)
\,,
\end{equation}
where we used $Q^2/W^2 \approx x$ for $W^2 \gg Q^2$. 
By this trick the effective dipole cross section got a $Q^2$-dependence. 

Of course, this is too simplistic and does not work since the integral in 
\eqref{F2modi} diverges at large $r$ due to \eqref{hbehav}. 
But we can modify the above 
argument slightly and replace $r$ by $r_\textrm{max}=z_0/Q$ 
only where $r$ is associated with the energy scale $W$. In this way 
we get 
\begin{equation}
\label{F2mod}
F_2^{\mathrm{mod}}(W,Q^2)= 
Q \int_0^\infty \ud r \,h(Qr,0) \,f\left(r, \frac{x}{z_0^2} \right)
\,.
\end{equation}
Now the integral is in general convergent and we obtain a dipole 
formula with a dipole-proton cross section depending on $r$ and $x$. 
That is, we made the replacement of scales \eqref{naivedipsize} 
\begin{equation}
\label{replacement}
\sum_q Q_q^2 \sigmadip^{(q)} (r,W) \:\: \longrightarrow \:\:
r^2 f\left(r, \frac{x}{C^2} \right) \equiv \hat{\sigma}^\mathrm{mod} (r,x)
\end{equation}
with $C = z_0=2.40$. But also other values of $C$ can be envisaged. 

We now show that also the replacement \eqref{replacement} is far from 
harmless and, in fact, modifies the structure function in an essential way. 
Basically this is due to the fact that the function $h(z,0)$ is very broad 
and is not well approximated by a delta function at $z_0$. 

\subsection{Substitution of scales via typical dipole sizes: examples}
\label{subsec:substex}

We first consider an example where a $Q^2$-dependence is introduced 
via the `typical dipole size' into a dipole cross section that originally 
depends only on $\rt$ and $W$ but not on $Q^2$.
Actually there are only few examples for dipole models of the latter kind 
in the literature. 
We choose here a slightly simplified version of a model proposed by 
Donnachie and Dosch in \cite{Donnachie:2001wt}. 
This DD model is based on Regge theory and includes exchanges 
of a soft and a hard pomeron. The intercepts of the soft and hard pomeron 
trajectories are denoted by $(1+\epsilon_s)$ and 
$(1+\epsilon_h)$, respectively. 
The dipole cross section that we want to consider is 
\begin{equation}
\label{DDsig}
\sigmadip^{(q)}_{\text{DD}}(r,W) = 
A_0 
r \left[ 1 - \exp \left( - \frac{r}{3.1 a} \right) \right] 
\left[ 
\theta (R_c - r) \left(\frac{r W}{R_c W_0}\right)^{2 \epsilon_h}
+ \theta (r -R_c) \left(\frac{W}{W_0}\right)^{2 \epsilon_s}
\right] \,.
\end{equation}
The parameter values are 
\begin{align}
\label{DDRpar}
\epsilon_h &= 0.42\,,&
A_0 &= 57.4 \, \mathrm{mb}/\mathrm{fm}\,,&
R_c &= 0.22 \, \mathrm{fm}\,,
\nonumber \\
\epsilon_s &= 0.08\,,&
W_0 &= 20\,\text{GeV} \,, &
a &= 0.346 \, \mathrm{fm}\,.&&
\end{align}
We take into account only light quarks ($u,d,s$) 
and set their masses to zero for simplicity. 
In appendix \ref{appB} we show how this simplified model is obtained 
from the original, more general approach of \cite{Donnachie:2001wt}. 
Depending on the kinematic parameters the simplifications lead to 
non-negligible deviations from the original model. 
But the simplified version is sufficient for the sake of our 
argument even if it describes the data only moderately well in 
some kinematic region.  
We should point out that both the simplified and the original model 
exhibit a rather strong rise of $F_2$ at very small $x$ and large 
$Q^2$ that originates from the assumed high intercept of the hard 
pomeron in this model. However, substantial deviations from the 
Golec-Biernat-W\"usthoff model (see below) occur only in regions 
in which no data are available. 

For the DD model we hence have 
\begin{equation}
\label{sigDDfDD}
\sigmadip_{\text{DD}} (r,W)= \sum_q Q_q^2 \sigmadip^{(q)}_{\text{DD}} (r,W)
= r^2 f_\mathrm{DD} \left(r,\frac{1}{r^2W^2}\right)
\,,
\end{equation}
where the sum is over the light flavours only and 
\begin{equation}
\label{DDdet}
\begin{split} 
f_\mathrm{DD} \left(r,\frac{1}{r^2W^2}\right)
= & \sum_q Q_q^2
\,A_0 \, \frac{1}{r}\, 
\left[ 1 - \exp \left( - \frac{r}{3.1 a} \right) \right] 
\\
& \times 
\left[ 
\theta (R_c - r) \left(\frac{r^2 W^2}{R_c^2 W_0^2}\right)^{\epsilon_h}
+ \theta (r -R_c) \left(\frac{r^2 W^2}{r^2 W_0^2}\right)^{\epsilon_s}
\right] \,.
\end{split}
\end{equation}

Making the replacement of scales \eqref{replacement} in \eqref{DDdet} gives 
\begin{equation}
\label{sigDDmod}
\hat{\sigma}_\mathrm{DD}^\mathrm{mod} (r,x) = 
r^2 f_\mathrm{DD} \left(r,\frac{x}{C^2}\right) 
\end{equation}
with $C=z_0=2.40$. 
As discussed above, this replacement changes the cross section only 
via terms which are sensitive to the external scale $W$, that is the 
terms in brackets with exponents $\epsilon_h$ and $\epsilon_s$ in this case. 
Figure \ref{fig:f2substDD} shows the effect of this substitution on the 
structure function $F_2$ obtained from these dipole cross sections 
(calculated only with light flavours). 
The solid curves show $F_2$ obtained using the (original) $W$-dependent 
dipole cross section $\sigmadip_{\text{DD}}(\rt,W)$, while the dashed curves 
show $F_2$ calculated using the $x$-dependent 
$\sigmadip_{\text{DD}}^{\text{mod}}(\rt,x)$. The results clearly deviate 
from each other. In particular, the $x$- and $Q^2$-dependences are heavily 
altered. 
This is also the case when one uses in the replacement of scales 
\eqref{naivedipsize} not $C=z_0$ but $C=2z_0=4.80$, 
corresponding to larger `typical dipole sizes'. $F_2$ obtained 
with this latter replacement is shown as the dotted curves in Figure \ref{fig:f2substDD}. 
\FIGURE[ht]{
\includegraphics[width=\textwidth]{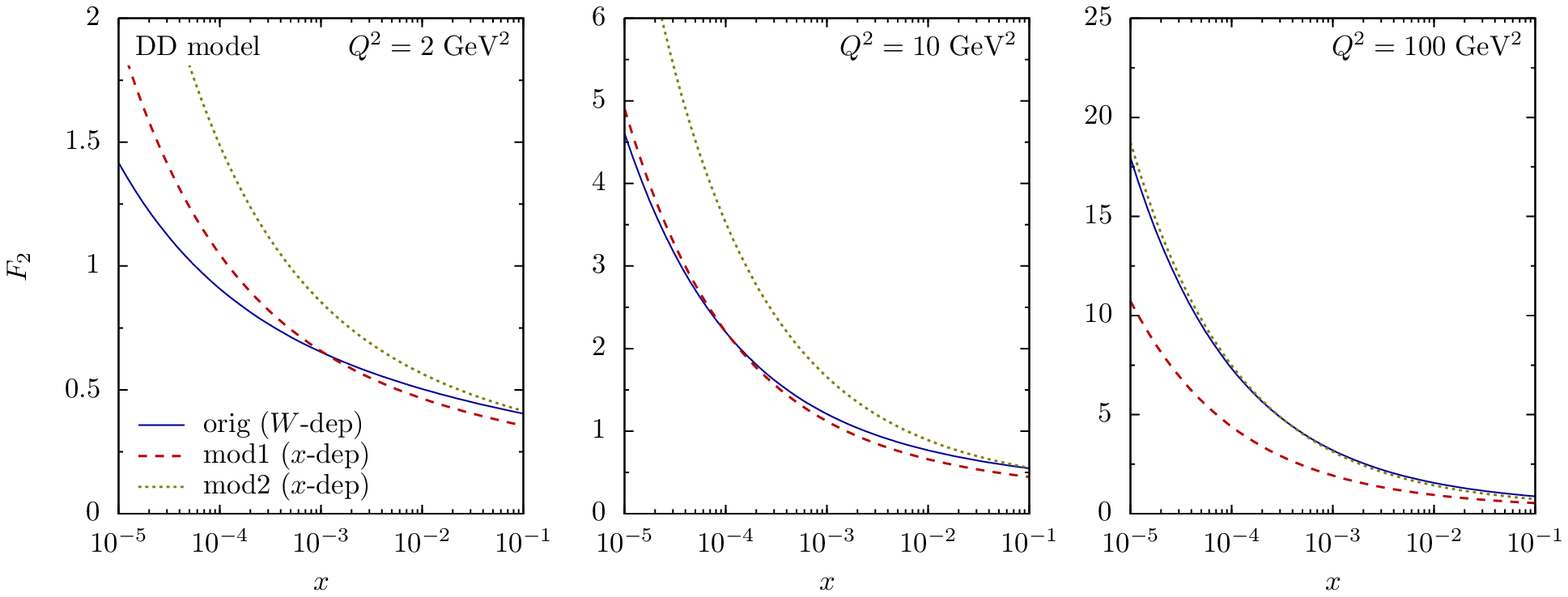}
\caption{
$F_2$ obtained from the simplified DD model for $\sigmadip$ 
and modified versions thereof, with 
energy dependencies via `typical dipole size' substitutions, for three 
values of $Q^2= 2,10,100\,\mbox{GeV}^2$. Bjorken-$x$ is varied 
and $Q^2$ is kept fixed for each plot.
Shown are the DD model with original $W$-dependence (solid curves)
and modifications of type \eqref{replacement} with $C=z_0=2.40$ (`mod1', dashed curves)
and $C=2z_0=4.80$ (`mod2', dotted curves).
\label{fig:f2substDD}
}
}

Let us now consider the reverse substitution of scales. That is, let 
us start from an $x$-dependent dipole cross section $\sigmadip (r,x)$ 
and make the reverse replacement of \eqref{replacement}, namely 
\begin{equation}
\label{revrepl}
\sigmadip (r,x) \equiv r^2 f\left(r, \frac{x}{C^2} \right) 
\:\: \longrightarrow \:\: 
r^2 f\left(r, \frac{1}{r^2 W^2} \right) \equiv \sigmadip^\mathrm{mod}(r,W)
\end{equation}
with $C=z_0=2.40$. 
As an example we consider the dipole model proposed by Golec-Biernat and 
W\"usthoff \cite{GolecBiernat:1998js}. 
This GBW model describes the $F_2$ data from HERA quite well. 
The dipole cross section of this model is given by 
\begin{equation}
\label{xsgbw}
\sigmadip_{\mathrm{GBW}}(r,x) = 
\sum_q Q_q^2 \sigmadip_{\mathrm{GBW}}^{(q)}(r,x) = 
\sum_q Q_q^2 \sigma_0 
\left[1-\exp\left(-\left(\frac{r}{2 R_0(x)}\right)^2\right)\right] \,.
\end{equation}
Accordingly, we have 
\begin{equation}
\label{fgbw}
f_{\mathrm{GBW}} \left(r, \frac{x}{C^2} \right) 
= \frac{1}{r^2} 
\sum_q Q_q^2 \sigma_0 
\left[1-\exp\left(-\left(\frac{r}{2 R_0(x)}\right)^2\right)\right] \,.
\end{equation}
For simplicity we consider also here only light ($u,d,s$) quarks, neglect 
their masses, and choose the following parameter set of \cite{GolecBiernat:1998js}: 
\begin{equation}
\label{GBWpar}
\begin{split}
R_0(x) = \left(\frac{x}{x_0}\right)^{\lambda/2} \mbox{GeV}^{-1} \, \,,\:\:\:\:\:
\sigma_0 = 23 \,\mbox{mb}, \:\:\:\:\:
\lambda=0.29\,,\:\:\:\:\:
x_0=3 \cdot 10^{-4} \,.
\end{split}
\end{equation}
The replacement \eqref{revrepl} gives us a modified dipole cross section 
$\sigmadip_\mathrm{GBW}^\mathrm{mod}(r,W)$. 
Figure \ref{fig:f2substGBW} 
shows the effect of this replacement on $F_2$. 
The curves for $F_2$ of the original GBW model (solid lines) are significantly
modified by the substitution \eqref{revrepl} (dashed lines), 
in particular the dependence on $Q^2$ and $x$ is altered. 
Again, this is also the case when using in the replacement of scales 
\eqref{naivedipsize} not $C=z_0$ but $C=2z_0=4.80$, 
corresponding to larger `typical dipole sizes', 
see the dotted lines in Figure \ref{fig:f2substGBW}. 
\FIGURE[ht]{
\includegraphics[width=\textwidth]{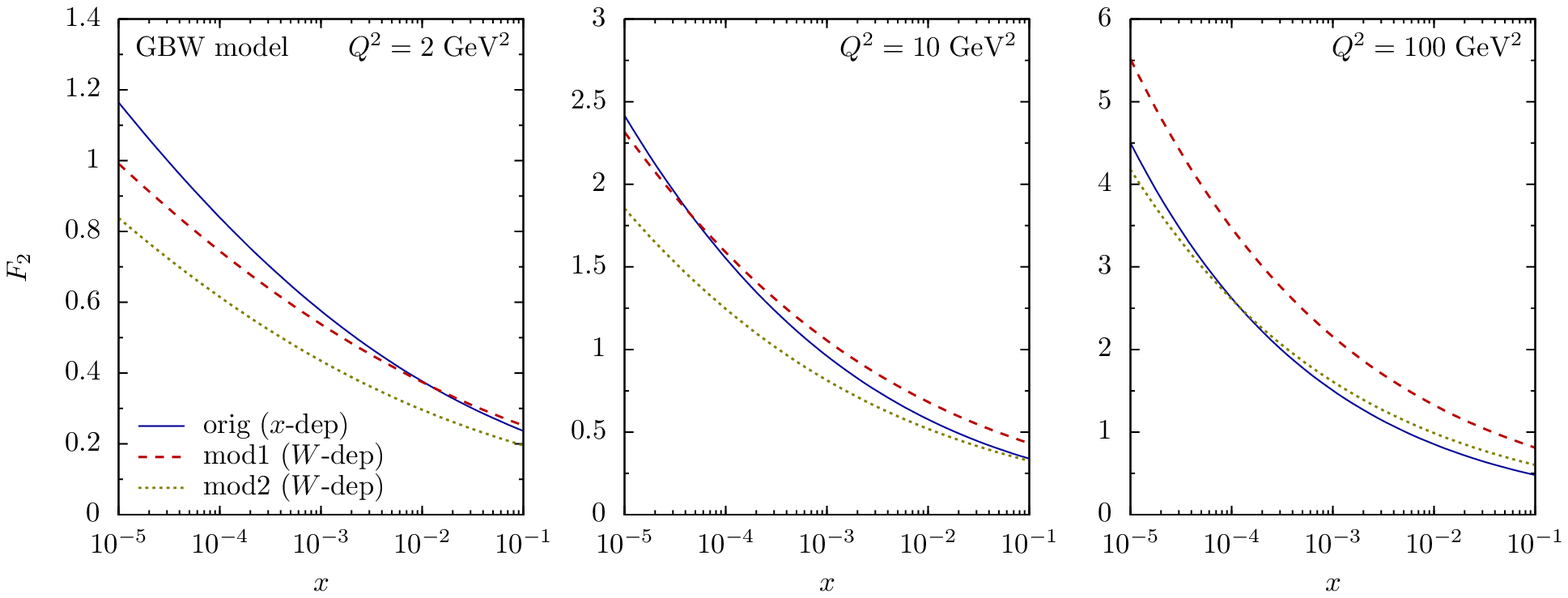}
\caption{
$F_2$ obtained from the GBW model for $\sigmadip$ and modified versions 
thereof, with energy dependencies via `typical dipole size' substitutions, for three 
values of $Q^2= 2,10,100\,\mbox{GeV}^2$. Bjorken-$x$ is varied 
and $Q^2$ is kept fixed for each plot.
Shown are the GBW model with original $x$-dependence (solid curves)
and modifications of type \eqref{revrepl} with $C=z_0=2.40$ (`mod1', dashed curves)
and $C=2z_0=4.80$ (`mod2', dotted curves).
\label{fig:f2substGBW}
}
}

In this section we have shown on two examples that the substitution 
of scales $r \leftrightarrow C/Q$ \eqref{naivedipsize} in the dipole 
cross section $\sigmadip$, done in either way, alters the structure 
function $F_2$ significantly. As could have been expected, the direction 
of the alteration is opposite in the two directions of performing 
the substitution. (Clearly, if one applies the substitution and its 
reverse subsequently on the same dipole cross section $\sigmadip$, 
the individual alterations have to cancel. The effect persists if one 
applies the two ways of substituting to different models which 
individually describe the data well.) The size of the alteration due 
to the substitution of scales is significant both quantitatively and 
qualitatively, see Figures \ref{fig:f2substDD} and \ref{fig:f2substGBW}. 

Actually, we can infer already from the results of 
\cite{Ewerz:2006an,Ewerz:2007md} that the choice of the energy 
variable in $\sigmadip$ is crucial at least at high $Q^2$. 
There it was shown that \emph{any} dipole cross section of the form 
$\sigmadip(\rt,W)$ fails to describe the HERA data for $Q^2 > 100$ to 
$200 \, \mbox{GeV}^2$. In contrast, the GBW model with 
$\sigmadip_{\text{GBW}}(\rt,x)$ provides a good fit to the data also at higher $Q^2$. 
From this we see that a dependence of $\sigmadip$ on $Q^2$ in addition to $W$ 
can certainly not be eliminated or introduced by an effective scale argument 
of the type \eqref{naivedipsize} in the regime of high $Q^2$ without drastic 
consequences. Let us, indeed, compare ratios of $F_2$ for different values of 
$Q^2$ for the models, original and modified, 
discussed above. An illustration of such ratios\footnote{Note that the curves 
are calculated from the models and shown here for a kinematic range that is 
slightly larger than that in which data from HERA are available. Corresponding 
curves restricted to the actual kinematic range where HERA data exist 
can be found in \cite{Ewerz:2006an,Ewerz:2007md}.} 
is given in \mfig \ref{fig:f2ratiosubst} for the example $W=150\,\mbox{GeV}$, 
where in addition the general bound (10) 
of \cite{Ewerz:2006an} is shown, which is valid for any $\sigmadip(\rt,W)$.
We see that, as expected, $\sigmadip_{\text{GBW}}^{\text{mod}}(\rt,W)$
as well as $\sigmadip_{\text{DD}}(\rt,W)$ respect the general bounds.
In contrast, the bounds are violated for both $x$-dependent cross
sections $\sigmadip_{\text{GBW}}(\rt,x)$ and 
$\sigmadip_{\text{DD}}^{\text{mod}}(\rt,x)$.%
\FIGURE[ht]{
\includegraphics[width=\textwidth]{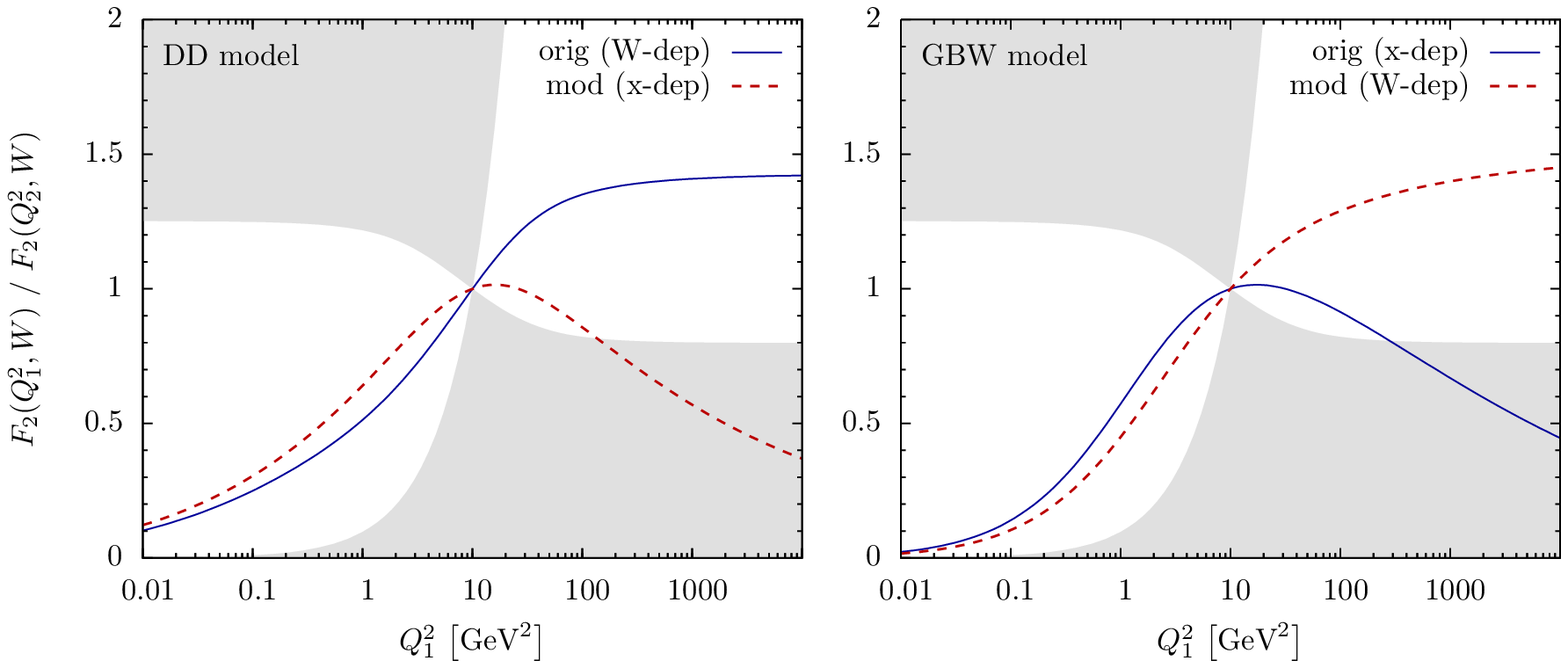}
\caption{\label{fig:f2ratiosubst}
Ratio $F_2(W,Q_1^2)/F_2(W,Q_2^2)$ for two dipole models and 
modified versions with altered energy dependencies via `typical dipole size' 
substitutions, for $W=150\,\mbox{GeV}$.  
Here, $Q_1^2$ is varied and $Q_2^2=10\,\mbox{GeV}^2$ is kept fixed.
Left plot for the (simplified) $W$-dependent Donnachie-Dosch model (solid curve)
and its modification \eqref{replacement} with $C=z_0=2.40$ (dashed curve).
Right plot for the original Golec-Biernat-W\"usthoff model (solid curve) 
and its modification using \eqref{revrepl} with $C=z_0=2.4$ (dashed curve).
Quark masses are set to zero in all cases.
The shaded area is excluded by the bound (10) of \cite{Ewerz:2006an}
for any $Q^2$-independent dipole cross section
$\sigmadip^{(q)}(r,W)$.
}
}

\section{The dipole formula as a convolution}
\label{sec:convolution}

In this section we consider again, for simplicity, only the light quarks
$u$, $d$, $s$ and neglect their masses.
Our starting point is the relation \eqref{Fwithhf} with $f$ inserted from 
\eqref{deff}. But now we leave the choice of energy variable open, 
\begin{equation} \label{f2xi}
 F_2(\xi,Q^2) = Q \int_0^\infty \ud r \,h(Q r, 0) \,\frac{1}{r^2} \,\hat\sigma(r,\xi) \,,
 \qquad
 \hat\sigma(r,\xi)=\sum_q Q_q^2 \hat\sigma^{(q)}(r,\xi) \,.
\end{equation}
We recall that the function $h(Qr,0)$ is obtained directly from the 
photon wave functions, see \eqref{defh}. 
Here we set $\xi=W$ if we want to consider a $W$-dependent dipole
cross section and $\xi=x$ for a $x$-dependent one.
Correspondingly, we consider $F_2$ as function of $\xi$ and $Q^2$.

Now we show that \eqref{f2xi} is a convolution formula.
For this we set, as in \eqref{zQr}, $Q r = z$ and define 
(recall that $z_0$ is defined as the position of the maximum of $h(z,0)$) 
\begin{align}
\label{taudef}
 \tau &= \ln (z/z_0) \,, \\
\label{kappadef}
 \kappa(\tau) &= z_0 h(z_0 e^\tau, 0) \,.
\end{align}
We have $0\le z < \infty$ and $-\infty < \tau < \infty$.
Corresponding to the behaviour of $h(z,0)$ for small and 
large $z$, see \eqref{hbehav}, we have
\begin{equation} \label{kappainfty}
\kappa(\tau) \propto \exp(\pm\tau) \qquad \quad \text{for } \tau \to \mp \infty \,.
\end{equation}
\FIGURE[ht]{
\includegraphics[width=0.6\textwidth]{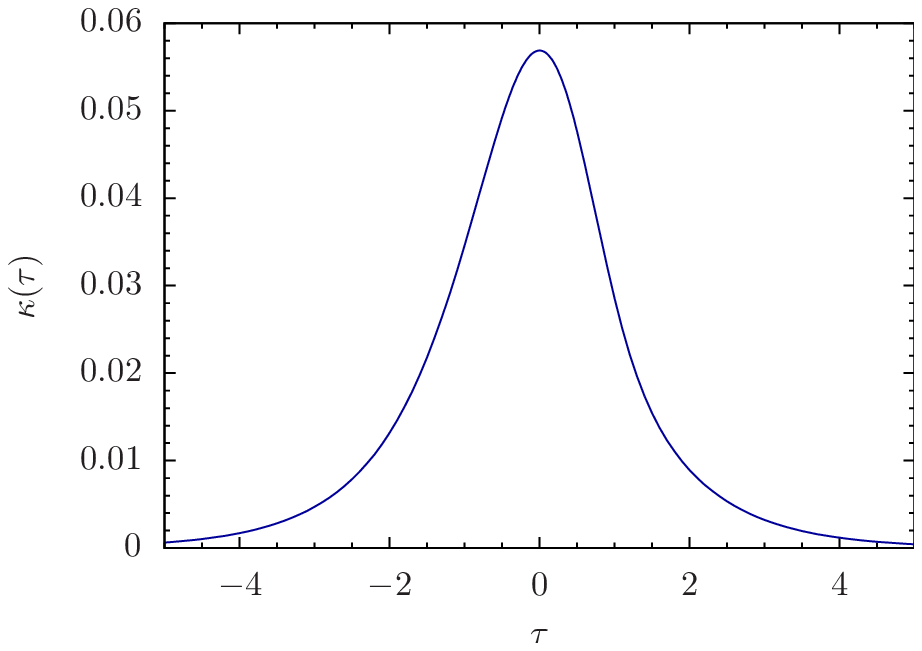}
\caption{\label{fig:kappa}  
The function $\kappa(\tau)$ defined in \eqref{kappadef}. 
}}
The function $\kappa(\tau)$ is shown in Figure \ref{fig:kappa}. 
The half width of this function is $\Delta\tau = 2.22$.
Now we choose $Q_0$ and $r_0$ such that
\begin{equation} \label{q0r0}
 Q_0 r_0 = z_0
\end{equation}
and define new variables, replacing $Q$ and $r$,
\begin{equation} \label{tdef}
 t = \ln(Q/Q_0)
\end{equation}
and
\begin{equation} \label{tpdef}
 t' = -\ln(r/r_0) \,.
\end{equation}
Furthermore, let us define a rescaled and reparametrised
`dipole cross section'
\begin{equation} \label{sdef}
 S(\xi,t') =
 \left. \frac{1}{r r_0} \,\hat\sigma(r,\xi)
 \right|_{r=r_0 \exp(-t')} 
\,.
\end{equation}
The dipole formula for massless quarks \eqref{f2xi} has now
the form of a convolution
\begin{equation} \label{f2emtconv}
 F_2(\xi, Q_0^2 e^{2t}) e^{-t} =
 \int_{-\infty}^{\infty}\ud  t' \,\kappa(t-t') S(\xi,t') \,.
\end{equation}
For dipole cross sections which are reasonably behaved for
small and large $r$ the integral in \eqref{f2emtconv} is convergent.
Indeed, we find
\begin{equation} \label{sinfty}
 S(\xi,t') \propto \exp(-t') \qquad \text{for } t'\to \infty,
 \quad\text{if }\hat\sigma(r,\xi) \propto r^2 \: \text{ for }\: r\to 0 
\end{equation}
and
\begin{equation} \label{sminfty}
 S(\xi,t') \propto \exp[(1-a)t'] \qquad \text{for } t'\to -\infty,
 \quad\text{if }\hat\sigma(r,\xi) \propto r^a \:\text{ for }\: r\to\infty \,.
\end{equation}
In the following we shall assume $a<1$. That is, we assume that
$\hat\sigma(r,\xi)$ increases more slowly than linearly with $r$
at large $r$.
In many models the dipole cross section saturates for large
$r$ which corresponds to $a=0$ in \eqref{sminfty}.
With these assumptions the function $S(\xi,t')$ decreases
exponentially for $t'\to\pm\infty$ and so does $\kappa(t-t')$.
Thus, the convolution formula \eqref{f2emtconv} is well convergent.

We assume now, for the sake of the argument, that for some
given $\xi$ the dipole formula \eqref{f2xi} is valid for all
$Q^2$, that is, \eqref{f2emtconv} holds for all $t$.
This allows us to draw some general conclusions:
\begin{enumerate}
\item[(i)]
The convolution formula \eqref{f2emtconv} by itself places really
no restrictions on $F_2$.
We can, in principle, invert the formula, using well known
techniques from the theory of integral transforms, 
and obtain $S(\xi,t')$ and the dipole cross section from $F_2$.
However, inversion is in general difficult and may constitute an 
ill-posed problem depending on the detailed structure of the 
functions involved. 
The physical restrictions enter in \eqref{f2emtconv} from the
requirement of a non-negative dipole cross section,
that is, from
\begin{equation} \label{spos}
 S(\xi,t') \geq 0 \qquad \text{for all } t' \,.
\end{equation}
Suppose now that we have indeed inverted the convolution
\eqref{f2emtconv} and found a non-negative $S(\xi,t')$. 
Then the dipole-proton cross section $\hat\sigma(r,\xi)$ will
appear as an integral over $Q^2$ of $F_2(\xi,Q^2)$,
respectively $\sigma_T(\xi,Q^2)+\sigma_L(\xi,Q^2)$,
using \eqref{f2dipsimple}.
Setting now $\xi=x$ we would find that $\hat\sigma(r,x)$ is
related to the $\gamma^\ast p$ cross sections at the same $x$
and all $Q^2$. This means that the dipole cross section at some 
fixed $r$ and $x$ would appear as (inverse) convolution of 
the structure function $F_2(x,Q^2)$ at the same fixed $x$ but 
varying $Q^2$ corresponding to very different hadronic 
final states $X$ in \eqref{processap}. 
Taken literally, the final states would range from the single proton
($Q^2=0$) to multi-particle final states with arbitrarily high 
invariant mass for $Q^2 \to \infty$. 
This looks not so plausible to us from the physics point of view.
On the other hand, setting $\xi=W$, only $\gamma^\ast p$ cross
sections for final states with the same c.\,m.\ energy will
appear in the integral for $\hat\sigma(r,W)$.
This appears much more plausible to us.
\item[(ii)]
The function $\kappa(t-t')$ in \eqref{f2emtconv} is rather broad
and decreases as $\exp(-t)$ for $t\to\infty$, see \eqref{kappainfty} 
and Figure \ref{fig:kappa}. Superposing $\kappa(t-t')$ with
non-negative weights $S(\xi,t')$ will lead to an even broader
function of $t$ on the l.h.s.\ of \eqref{f2emtconv}.
That is, the decrease of $F_2(\xi, Q_0^2 e^{2t}) \exp(-t)$
can then not be faster than $\exp(-t)$ for $t\to\infty$.
This means that the dipole formula will be in difficulties if,
for large $Q^2$, $F_2(\xi,Q^2)$ \emph{decreases} with $Q^2$
at fixed $\xi$ but will work if $F_2(\xi,Q^2)$ \emph{increases}
with $Q^2$ at fixed $\xi$.
From this simple observation we can already see why,
phenomenologically, dipole models with the choice $\xi=x$ work
up to higher $Q^2$ than models where the choice is $\xi=W$.
For fixed small $x$ the structure function $F_2(x,Q^2)$ 
increases with $Q^2$ in the HERA regime, see left panel in 
Figure \ref{fig:f2dep}. For fixed $W$, in contrast, $F_2(W,Q^2)$ 
first increases, but eventually decreases with increasing $Q^2$, 
see right panel in Figure \ref{fig:f2dep}. 
\end{enumerate}%
\FIGURE{
\includegraphics[width=\textwidth]{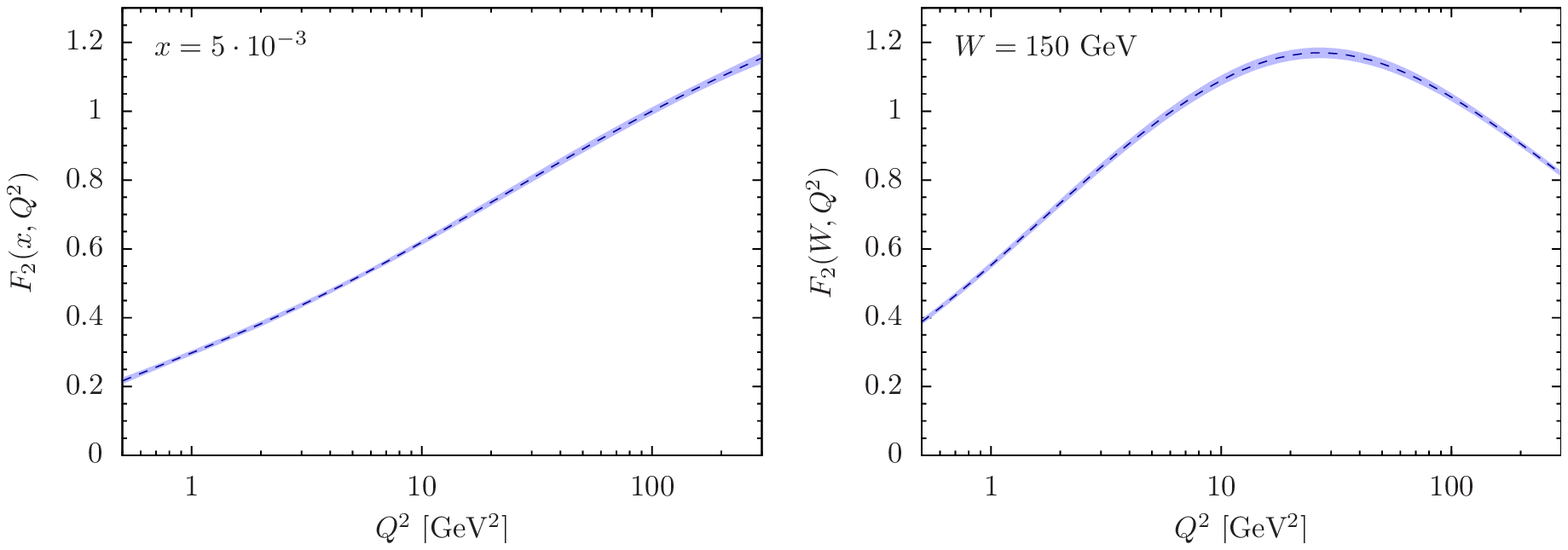}
\caption{\label{fig:f2dep}
$Q^2$-dependence of the structure function $F_2$. On the left for 
fixed $x=5 \cdot 10^{-3}$, on the right for fixed $W=150\,\mbox{GeV}$. 
$F_2$ shown here is obtained from the parametrisation of 
\cite{Del Debbio:2004qj}, the band indicates a $1 \sigma$-uncertainty. 
}
}

We shall now illustrate the convolution formula \eqref{f2emtconv} for
the example of the GBW model where we have, of course, the
choice $\xi=x$, see \eqref{xsgbw} and \eqref{GBWpar}. 
Furthermore we set 
\begin{equation} \label{q0r0nums}
 Q_0 = \sqrt{10}\mbox{ GeV}\,, \qquad
 r_0 = z_0/Q_0 = 0.24 \mbox{ GeV}^{-1} \: \widehat{=} \:
      0.047 \mbox{ fm} \,.
\end{equation}
In Figure \ref{fig:skappa} we show for $x=10^{-3}$ the original function
$S_\mathrm{GBW}$, suitably normalised, 
\begin{equation} \label{sintkap}
 ( S \int \kappa )_\mathrm{GBW}(x,t) :=
 S_\mathrm{GBW}(x,t) \int_{-\infty}^{\infty} \ud \tau\, \kappa(\tau)
 = \left. \frac{0.15}{r r_0} \, \hat\sigma_\mathrm{GBW}(r,x)
   \right|_{r=r_0 \exp(-t)} \,,
\end{equation}
as well as the result of the convolution,
\begin{equation} \label{sconvkap}
 (S \ast \kappa)_\mathrm{GBW}(x,t) :=
 \int_{-\infty}^{\infty} \ud t'  \,\kappa(t-t') S_\mathrm{GBW}(x,t')
 = \left. F_{2\,\mathrm{GBW}}(x,Q^2) \,\frac{Q_0}{Q}
   \right|_{Q=Q_0 \exp(t)} \,.
\end{equation}
The effect of the broadening due to the convolution is clearly
visible in Figure \ref{fig:skappa}. We also see that within a factor
of 2 we have very roughly 
\begin{equation} \label{sconvkapcent}
 (S \ast \kappa)_\mathrm{GBW}(x,t) \approx
 (S \int \kappa)_\mathrm{GBW}(x,t)
\qquad \text{for } -3 \lesssim t \lesssim 1 \,. 
\end{equation}
This relation actually holds for a wide range in $x$, with the $t$-range 
varying somewhat with $x$. 
Translating this back to $r$ and $Q$ we find that within a factor
of 2 we have 
\begin{equation} \label{f2qcent}
 F_{2\,\mathrm{GBW}}(x,Q^2) \frac{Q_0}{Q} \approx
 \left. \frac{0.15}{r r_0} \, \hat\sigma_\mathrm{GBW}(r,x)
 \right|_{r = z_0/Q}
 \qquad \text{for }\: 0.2\,\mbox{GeV} \lesssim Q \lesssim 8 \,\mbox{GeV} \,.
\end{equation}
\FIGURE{
\includegraphics[width=0.6\textwidth]{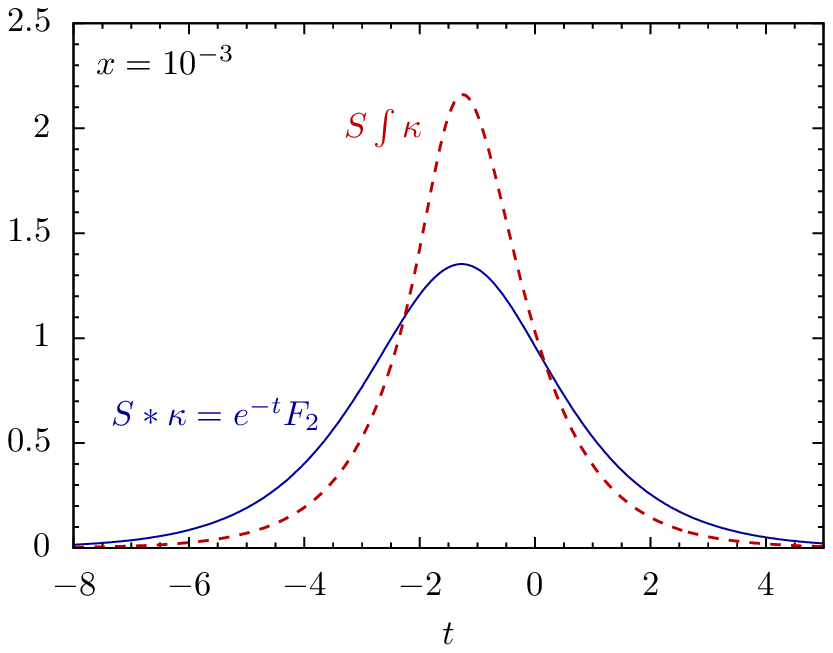}
\caption{\label{fig:skappa}
The functions \eqref{sintkap} (red, dashed) and 
\eqref{sconvkap} (blue, solid) in the GBW model for $x=10^{-3}$.
}}

But for large and for very small $Q^2$ a relation of the form \eqref{sconvkapcent}
respectively \eqref{f2qcent} does \emph{not} hold at all.
In the left panel of Figure \ref{fig:skappaet} we show the functions 
\eqref{sintkap} and \eqref{sconvkap} of Figure \ref{fig:skappa} 
but multiplied by $\exp(t)$.
This gives a clear indication of what happens for large $t$,
that is, large $Q^2$ and small $r$.
We see from the Figure that, for $t \gtrsim -1$, 
$F_{2\,\mathrm{GBW}}=e^t(S\ast\kappa)_\mathrm{GBW}$ and
$e^t(S\int\kappa)_\mathrm{GBW} =
0.15 \, r^{-2}\hat\sigma_\mathrm{GBW}|_{r=r_0 \exp(-t)}$
have a completely different behaviour. 
$F_{2\,\mathrm{GBW}}$ rises linearly with $t$, while 
$\hat\sigma_\mathrm{GBW}/r^2$ goes to a constant for
$t\to\infty$, that is $r\to 0$.
The latter corresponds to the assumption of colour transparency
in the GBW model.
Note that the derivative of a linear function of $t$
gives a constant.
We shall show below that quite generally we expect indeed 
a derivative relation between $F_2(\xi,Q^2)$ and
$r^{-2}\hat\sigma(r,\xi)$ for large $Q^2$. 
\FIGURE{
\includegraphics[height=5.4cm]{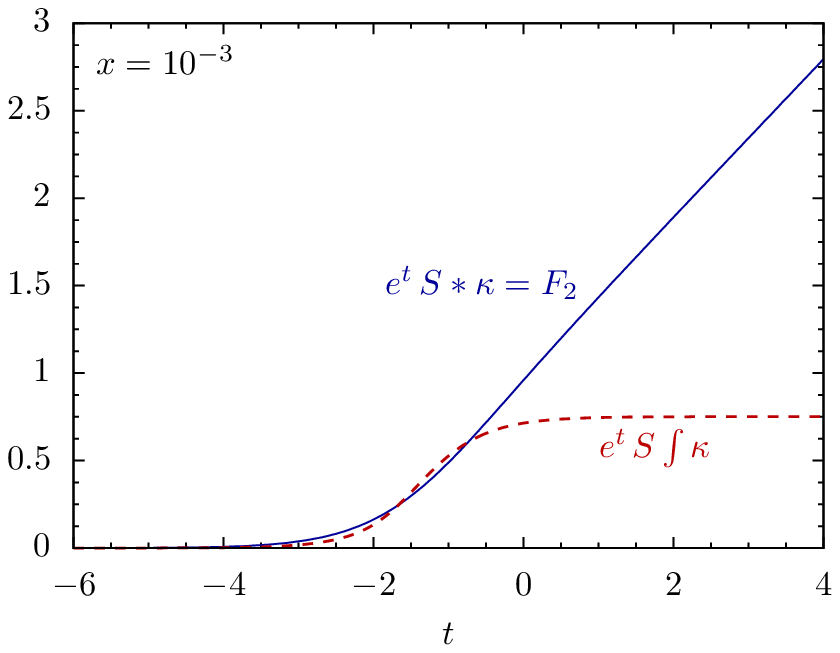}
\hspace*{0.5cm}
\includegraphics[height=5.4cm]{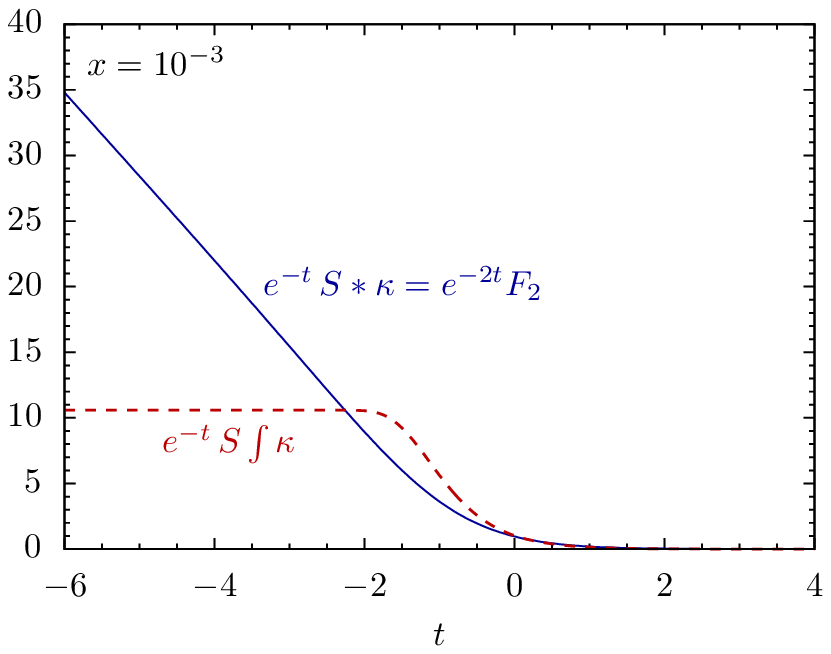}
\caption{\label{fig:skappaet}
The two functions of Figure \ref{fig:skappa}, but now 
multiplied by $\exp(t)$ (left panel) and $\exp(-t)$ (right panel). 
}}

In the right panel of Figure \ref{fig:skappaet} we show again 
the functions \eqref{sintkap} and \eqref{sconvkap} of Figure \ref{fig:skappa} 
but now multiplied by $\exp(-t)$.
In this way we see clearly what happens for $t\to -\infty$,
that is, for very small $Q^2$ and large $r$.
For $t\to -\infty$ the function $e^{-t} S \ast \kappa = e^{-2t} F_2$ 
behaves as $(-t)$ whereas $e^{-t} S \int \kappa$ is constant, 
suggesting again a derivative relation between them. 
Indeed, we shall show below that on general grounds we expect for
$t\to -\infty$ such a derivative relation between $\sigma_T+\sigma_L$
and $\hat\sigma$ to hold. 

\boldmath
\subsection{The large-$Q^2$ regime}
\unboldmath

In this subsection we give a general discussion of the convolution
formula \eqref{f2emtconv} for large $t$ corresponding to large $Q^2$
and small $r$.
We rewrite \eqref{f2emtconv} in the form
\begin{equation} \label{f2conv}
 F_2(\xi,Q_0^2 e^{2t}) =
 \int_{-\infty}^{\infty}\ud t' \,\kappa(t-t') \exp(t-t') e^{t'} S(\xi,t')
\end{equation}
where
\begin{equation} \label{set}
 e^{t'} S(\xi,t') =
 \left. \frac{1}{r^2}\hat\sigma(r,\xi) \right|_{r=r_0 \exp(-t')}
\,.
\end{equation}
\FIGURE{
\includegraphics[width=0.45\textwidth]{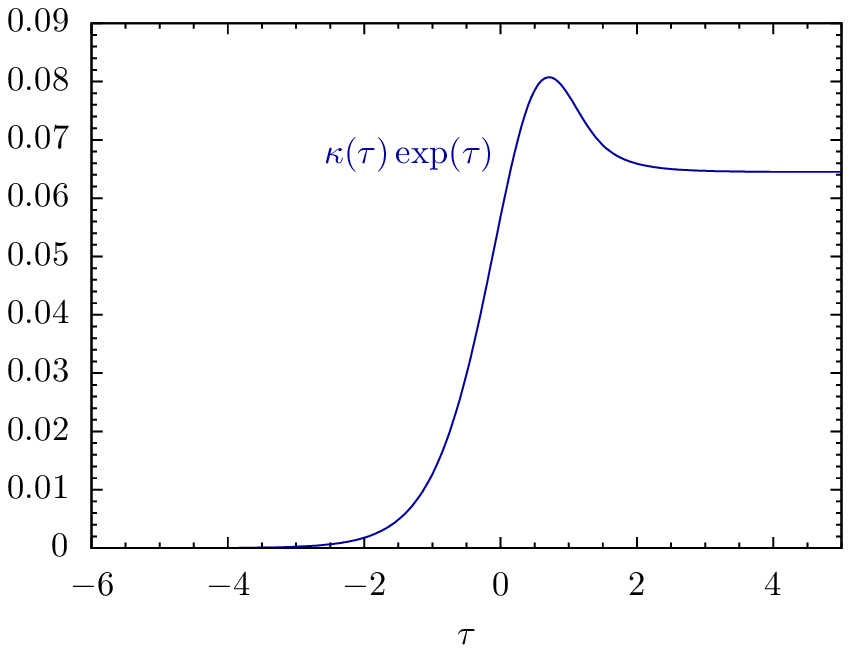}
\hspace*{.8cm}
\includegraphics[width=0.45\textwidth]{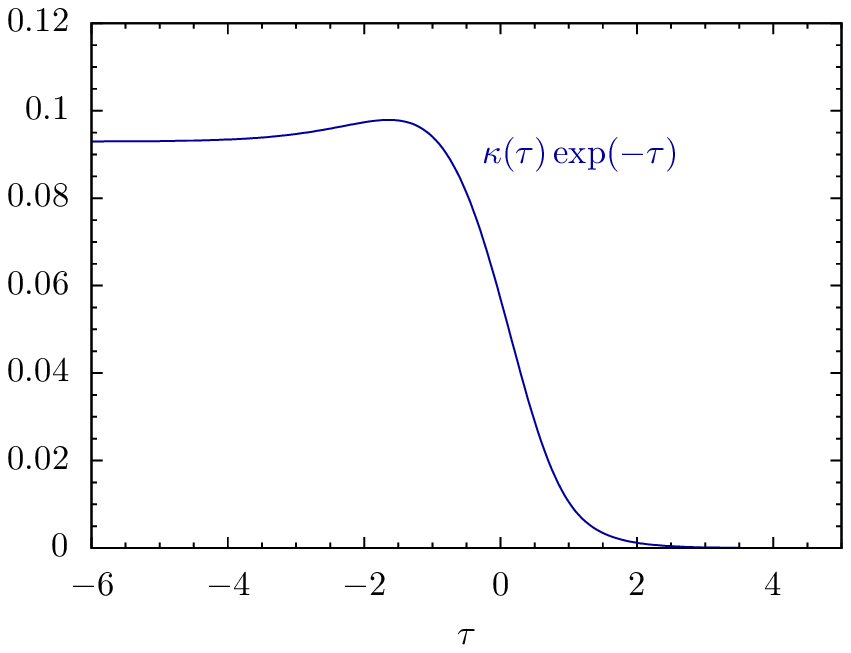}
\caption{\label{fig:kappaetau}  
The functions $\kappa(\tau) \exp(\tau)$ and $\kappa(\tau) \exp(-\tau)$. 
}}

The function $\kappa(\tau) \exp(\tau)$ is shown in the left panel of 
Figure \ref{fig:kappaetau}. Very roughly it resembles a step function.
Suppose now that $\hat\sigma(r,\xi)/r^2$ is slowly varying for 
$r\to 0$ corresponding to $\exp(t') S(\xi,t')$ being slowly varying
for $t'\to\infty$.
In \eqref{f2conv} this function is folded with $\kappa(t-t')\exp(t-t')$
which resembles a step function.
The folding produces, therefore, in essence the primitive of
$\exp(t) S(\xi,t)$.
That is, we expect for large $t$ the following relation to hold 
\begin{equation} \label{setderiv}
 c_{\infty} e^t S(\xi,t) \approx
 \frac{\partial}{\partial t} \, F_2(\xi, Q_0^2 e^{2t})
 \qquad (t \gg 1) \,,
\end{equation}
respectively
\begin{equation} \label{xhr2deriv}
 c_\infty \frac{1}{r^2} \,\hat\sigma(r,\xi) \approx
 \left. 2 \, Q^2 \frac{\partial}{\partial Q^2} \, F_2(\xi,Q^2)
 \right|_{Q^2 = (z_0/r)^2}
 \qquad (Q^2 \gg 74\,\mbox{GeV}) \,,
\end{equation}
with
\begin{equation}
c_\infty := 2 \pi (\kappa(\tau) \left. \exp(\tau)) \right|_{\tau \to \infty}
             = \frac{2}{\pi^3}
             = 0.065 \,.
\end{equation}
This is illustrated for the GBW model in the left panel of 
Figure \ref{fig:derrel} for the choice $x=10^{-3}$. 
Here the relation \eqref{setderiv} respectively
\eqref{xhr2deriv} is indeed well confirmed for 
$t>0$ corresponding to $Q^2 > 10~\mbox{GeV}^2$.
\FIGURE{
\includegraphics[height=5.4cm]{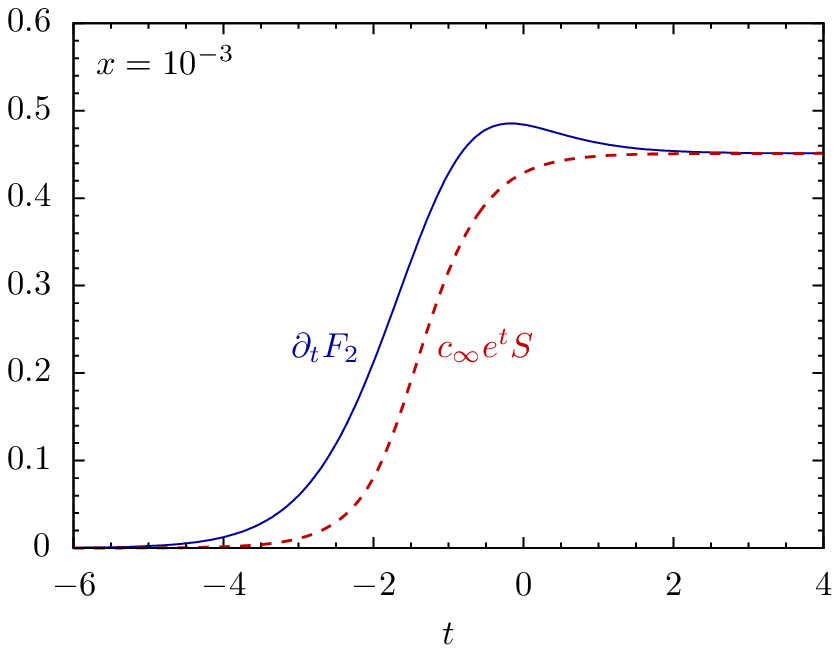}
\hspace*{0.5cm}
\includegraphics[height=5.4cm]{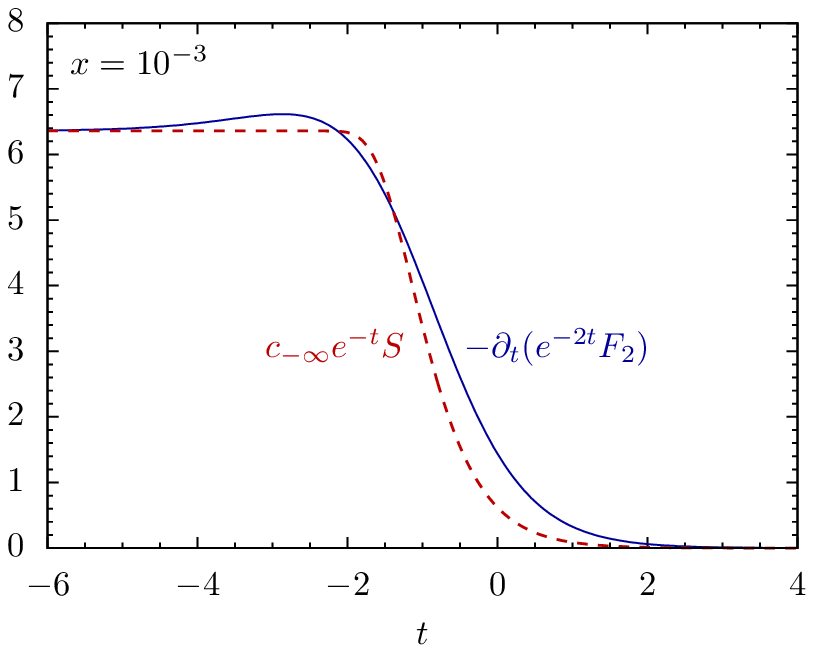}
\caption{\label{fig:derrel}  
Numerical test of the derivative relations \eqref{setderiv} respectively 
\eqref{xhr2deriv} (left panel) and \eqref{semtderiv} respectively 
\eqref{xhderiv} (right panel) for the GBW model for $x=10^{-3}$. 
}}

From \eqref{xhr2deriv} we see that the dipole formula will work well
phenomenologically at high $Q^2$ if we choose an energy variable
$\xi$ such that $F_2(\xi,Q^2)$ increases with $Q^2$ at fixed $\xi$
since then the cross section $\hat\sigma$ can be positive.
This is the case for the choice $\xi=x$ for small $x$ but,
of course, many other choices of $\xi$ would also have this property.
On the other hand, $F_2(W,Q^2)$ decreases for fixed $W$ at
large enough $Q^2$, see figure \ref{fig:f2dep}.
Therefore, it is clear that dipole models with the choice
$\hat\sigma(r,W)$ will only be able to describe the structure
function $F_2(W,Q^2)$ over a more limited range of $Q^2$ values.
But this limitation at high $Q^2$ is as it \emph{should be} 
and is physically reasonable as explained 
in \cite{Ewerz:2006vd,Ewerz:2006an,Ewerz:2007md}. 
Therefore, this should by no means be used as argument against
the choice of the energy variable $W$.

\boldmath
\subsection{The small-$Q^2$ regime}
\unboldmath

To study the dipole formula in the small-$Q^2$ regime, that is
for $t\to -\infty$, we rewrite \eqref{f2emtconv} in the following form
\begin{equation} \label{f2em2tconv}
 F_2(\xi,Q_0^2 e^{2t})e^{-2t} =
 \int_{-\infty}^{\infty} \ud t' \,\kappa(t-t') \exp[-(t-t')]
 \exp(-t') S(\xi,t')
\,.
\end{equation}
The function $\kappa(\tau)\exp(-\tau)$ resembles very roughly
a downward step function, see the right panel of Figure \ref{fig:kappaetau}.
Furthermore we have 
\begin{equation} \label{semt}
 \exp(-t) S(\xi,t) =
 \left. \frac{1}{r_0^2} \, \hat\sigma(r,\xi)\right|_{r=r_0 \exp(-t)}
\,,
\end{equation}
and we study here the limit $t\to -\infty$, that is,
the limit of large $r$.
Let us assume now that $\hat\sigma(r,\xi)$ is a smooth,
only slowly varying, function of $r$ for $r\to\infty$.
This is, for instance, the case for dipole models where the
dipole-proton cross section saturates for $r\to\infty$,
\begin{equation} \label{xhsat}
 \hat\sigma(r,\xi) \rightarrow \sigma_0
 \qquad \text{for }\: r\to\infty \,.
\end{equation}
Now we can apply the same type of argument as we used for
the large-$Q^2$ regime.
In \eqref{f2em2tconv} the slowly varying function $\exp(-t')S(\xi,t')$
is, for $t\to -\infty$, folded with a function resembling
a downward step function.
Thus, we get an approximate derivative relation
\begin{equation} \label{semtderiv}
 c_{-\infty} \exp(-t) S(\xi,t) \approx
 -\frac{\partial}{\partial t}
 \left[ F_2(\xi,Q_0^2 e^{2t}) e^{-2t}) \right] \,,
\end{equation}
or, put differently,
\begin{equation} \label{xhderiv}
 c_{-\infty} \frac{1}{r_0^2} \hat\sigma(r,\xi) \approx
 \frac{Q_0^2}{4\pi^2\alpha_\mathrm{em}}
 \left. (-2) \,Q^2 \frac{\partial}{\partial Q^2} \,
  [\sigma_T(\xi,Q^2) + \sigma_L(\xi,Q^2)]
 \right|_{Q^2=(z_0/r)^2} \,,
\end{equation}
where 
\begin{equation}
c_{-\infty} := 2 \pi (\kappa(\tau) \left. \exp(-\tau))\right|_{\tau \to -\infty}
             = \frac{z_0^2}{2 \pi^3} = 0.093 \,.
\end{equation}
In the right panel of Figure \ref{fig:derrel} we show for the GBW model 
the functions of the l.h.s.\ and r.h.s.\ of \eqref{semtderiv}. 
We see that there is indeed approximate equality of the two functions 
for $t< -2$. Translated into cross section implies \eqref{xhderiv} 
to hold for $Q^2 < 0.2\,\mbox{GeV}^2$. These are extremely small 
$Q^2$, and the actual relation is therefore more of academic than 
of phenomenological interest. Nevertheless, we find it useful to discuss 
its meaning. As we will show now, \eqref{xhderiv} gives interesting 
insight concerning the approximations underlying the dipole picture. 

The result \eqref{xhderiv} is somewhat surprising. It means the
following.
The dipole model with energy variable $\xi$ can only work
down to very small $Q^2$ with non-negative cross section
$\hat\sigma(r,\xi)$ if $\sigma_T(\xi,Q^2) + \sigma_L(\xi,Q^2)$ 
increases with decreasing $Q^2$ for small $Q^2$.
A saturating dipole cross section
\begin{equation} \label{xhsatnz}
 \hat\sigma(r,\xi) \rightarrow \sigma_0 \neq 0
 \qquad \text{for } \: r\to\infty
\end{equation}
implies a \emph{logarithmic increase} of $\sigma_T+\sigma_L$:
\begin{equation} \label{xtlzero}
 \sigma_T(\xi,Q^2)+\sigma_L(\xi,Q^2) \propto \ln(Q_0^2/Q^2)
 \qquad \text{for } \: Q^2\to 0 \,.
\end{equation}
Clearly, the behaviour \eqref{xtlzero} is unacceptable physically.
For the case $\xi=W$ we know that gauge invariance requires
the behaviour
\begin{equation} \label{xtlwzero}
 \sigma_T(W,Q^2) + \sigma_L(W,Q^2) \rightarrow \text{const.}
 \qquad \text{for } \: Q^2 \to 0\,.
\end{equation}
Indeed, a realistic behaviour for $Q^2 \to 0$ is 
\begin{equation}
\label{realbehav}
\sigma_T (W,Q^2) + \sigma_L(W,Q^2) = 
\sigma_{\gamma p}(W) \left( 1 - \frac{Q^2}{m^2_{\rm eff}} + \mathcal{O}(Q^4) \right) \,,
\end{equation}
where $\sigma_{\gamma p}(W)$ is the real-photon-proton cross section and 
$m^2_{\rm eff} > 0$ is a constant of dimension mass squared. 
Inserting \eqref{realbehav} in \eqref{xhderiv} leads to 
\begin{equation} 
\label{4.31b}
\hat{\sigma} (r,W) \approx \frac{1}{c_{- \infty}} \,
\frac{z_0^4 \sigma_{\gamma p}(W)}{2 \pi^2 \alpha_{\rm em}} \frac{1}{r^2 m^2_{\rm eff}} 
\left( 1 + \mathcal{O} \left(\frac{r_0^2}{r^2} \right) \right)
\end{equation}
for $r \to \infty$. 

Thus, we have two options. We can either assume that the naive dipole picture 
with the standard perturbative photon wave functions holds down to very small 
$Q^2$ values. Then, the above arguments force us to give up the saturation 
hypothesis \eqref{xhsatnz} for the dipole-proton cross section $\hat{\sigma}(r,W)$ 
which should then instead vanish as $1/r^2$ for $r\to \infty$, see \eqref{4.31b}.  
We think, however, that the more likely resolution of the above puzzle is that the 
naive dipole picture must be modified for small $Q^2$. 
But we know already from the discussions in 
\cite{Ewerz:2006vd,Ewerz:2006an,Ewerz:2007md} 
that the naive dipole picture is expected to break down
for small $Q^2$, where the estimate was $Q^2 \lesssim 2~\mbox{GeV}^2$.
Thus, the discussion above gives further evidence for this breakdown.

In \cite{Ewerz:2006vd} a general analysis of the modifications and additions 
to the photon wave functions and the dipole cross section to be expected 
for small $Q^2$ was given. We believe that these corrections are relevant in 
the context of the discussion above. On a more phenomenological level, 
one can also cure the problem by introducing modifications of the photon 
wave functions, as done for instance in \cite{Dosch:1997nw}, and get then 
a reconciliation of \eqref{xhsatnz} and \eqref{xtlwzero}. 
The implementation of vector meson dominance at small $Q^2$ has 
similar effects with regard to the problem pointed out here. 

\section{Conclusions}
\label{sec:conclusions}

The dipole model is widely used in the analysis of deep inelastic 
scattering data, but also in other processes. Often one attempts to 
extract subtle effects from the data with the help of the dipole picture, 
for example the presence or absence of saturation effects. Many of 
these studies use Bjorken-$x$ as the energy variable in the 
dipole cross section $\sigmadip$. However, general considerations based 
on quantum field theory lead to the conclusion that the correct 
energy variable in $\sigmadip$ is $W$. In particular, 
the dipole cross section has to be independent of the photon 
virtuality $Q^2$. This issue of the energy variable in $\sigmadip$ 
has not been taken very seriously so far, since the integral over 
dipole sizes is thought to be dominated by dipoles of a typical 
size $r \sim 1/Q$. Assuming the strong dominance of a typical 
dipole size one can freely change from one energy variable to 
another, in particular from $W$ to $x$ and vice versa. 
We have shown here that this simple picture is misleading. 
In particular, we have shown that the corresponding change of 
energy variable has a large effect on the resulting structure function 
for any given model of the dipole cross section. Numerically, the 
resulting effect is sizeable and can easily exceed the spread among 
different models for $\sigmadip$ suggested in the literature. 
For a reliable study of subtle effects in the data the correct choice of 
energy variable hence appears much more important than has been 
previously assumed. 

Apart from the numerical consequences of choosing energy variables 
other than $W$ in the dipole cross section there is also an important 
conceptual reason for choosing $W$. With the variables used 
in \eqref{sigmatdip} that formula is an actual factorisation: 
the variables of the l.h.s., $Q^2$ and $W$, are separated and occur in 
two different factors in the integrand on the r.h.s. If the second factor 
in the integrand on the r.h.s.\ would still contain $x$ and hence $Q^2$, 
the dipole formula would not constitute a factorisation. This would 
spoil the familiar (and correct) picture of photon-proton scattering 
as a two-step process: the photon fluctuating into a quantum-mechanical 
superposition of dipoles of all possible sizes and the subsequent 
dipole-proton scattering. The photon virtuality $Q^2$ determines the 
probability distribution of dipole sizes, but a dipole of given size does 
not inherit any information on $Q^2$. Its subsequent scattering on 
the proton hence cannot depend on $Q^2$. If, as is usually done in the 
dipole picture for DIS, the second step is interpreted as the cross section 
of asymptotic dipole states on the proton, the use of an energy variable 
involving $Q^2$ in the dipole cross section would violate the rules of 
quantum field theory. 

In the second part of the paper we have shown that the dipole formula 
can be written such that the structure function $F_2$ is represented 
by a convolution of the (suitably rescaled) dipole cross section with 
a known function originating from the photon wave functions. In principle, 
this convolution could even be invertible, albeit with some caveats. 
While a precise determination of the dipole cross section from $F_2$ 
based on inversion of the convolution appears very difficult we have 
found that the asymptotic behaviour of the dipole cross section 
$\sigmadip(r,\xi)$ for small and large $r$ can be extracted from $F_2$ 
(even for any given choice of energy variable $\xi$). In particular, 
we have found derivative relations between $F_2$ and the dipole cross 
section for both large and very small $Q^2$. We expect that these 
findings can be used to construct improved models for the dipole cross 
section. 

Further, we have been able to explain why in general dipole models 
using the energy variable $x$ are better suited for fitting $F_2$ 
in a wide range of kinematic parameters than models using the 
correct variable $W$. Based on the representation of the dipole formula 
as a convolution this can be 
directly traced back to the sign of the scaling violations that occur 
for increasing $Q^2$ at fixed small $x$. In our opinion that does not 
disfavour the choice of $W$ as energy variable in the dipole cross section. 
On the contrary, the dipole model clearly contains approximations implying 
already a priori a limited kinematic range for its applicability. 
As shown in \cite{Ewerz:2006vd,Ewerz:2006an,Ewerz:2007md} 
the choice of $W$ as energy variable leads to such limits which are physically 
reasonable. 

\section*{Acknowledgements}

We would like to thank J.-P.\ Blaizot, H.\,G.\ Dosch, 
K.\ Golec-Biernat, and A.\ Shoshi for helpful discussions. 
C.\,E.\ was supported by the Alliance Program of the
Helmholtz Association (HA216/EMMI). 
A.\,v.\,M.\ was supported by the Schweizer Nationalfonds.
This work was supported by the Deutsche Forschungsgemeinschaft, 
project number NA 296/4-1. 

\begin{appendix}

\section{Integrated photon densities}
\label{appA}

In this appendix we derive the asymptotic behaviour of the 
(over longitudinal momentum fraction $\alpha$) integrated photon densities
$w_{T,L}^{(q)}(r,Q^2)$, defined in \eqref{denst} and \eqref{densl}, 
at small and large distances $r$. In the main text these limits are needed 
only for massless quarks. For completeness we give them for massive 
quarks as well. 

At small distances, $r Q \ll 1$ and $r m_q \ll 1$,
we find from \eqref{sumpsitdens} and \eqref{sumpsildens}
\begin{align}
\label{wtr0}
w_T^{(q)}(r,Q^2) &\,\xrightarrow{r \rightarrow 0}\,
  \frac{N_c\alpha_\mathrm{em} Q_q^2}{2\pi^2} \,Q^2
    \frac{2}{3}\,\frac{1}{(Q r)^2}\,,\\
\label{wlr0}
w_L^{(q)}(r,Q^2) &\,\xrightarrow{r \rightarrow 0}\,
  \frac{2 N_c\alpha_\mathrm{em} Q_q^2}{\pi^2} \,Q^2 \frac{1}{30}\,\ln^2 (Q r)\,.
\end{align}
Note that these leading terms are independent of $m_q$.

For the limits at large distances we consider massless quarks and
massive quarks separately.
For massless quarks, $m_q=0$, we find without any approximation
\begin{align}
w_T^{(q)}(r,Q^2) &= \frac{N_c\alpha_\mathrm{em} Q_q^2}{2\pi^2} \,Q^2
  \frac{2\pi}{(Q r)^6} \Bigg[ 
  2\, G^{3,1}_{2,4}\left( \frac{1}{4}(Q r)^2 \left\vert
    {\begin{array}{l} 2, 7/2 \\ 2, 3, 4, 3/2 \end{array}} 
\right.
    \right)
\notag\\
&\qquad\qquad\qquad\qquad\qquad
 - G^{3,1}_{2,4}\left( \frac{1}{4}(Q r)^2 \left\vert
    {\begin{array}{l} 1, 7/2 \\ 2, 3, 4, 1/2 \end{array}} \right.
    \right)
  \Bigg],\\
w_L^{(q)}(r,Q^2) &= \frac{2 N_c\alpha_\mathrm{em} Q_q^2}{\pi^2} \,Q^2
  \frac{\pi}{(Q r)^6} \,
    G^{3,1}_{2,4}\left( \frac{1}{4}(Q r)^2 \left\vert
    {\begin{array}{l} 1, 7/2 \\ 3,3,3,1/2 \end{array}} \right.
    \right)
\end{align}
with Meijer's $G$-function.
From this we get at large distances, $r Q \gg 1$:
\begin{align}
\label{wtrinfzero}
w_T^{(q)}(r,Q^2) &\,\xrightarrow{r \rightarrow \infty}\,
  \frac{N_c\alpha_\mathrm{em} Q_q^2}{2\pi^2} \,Q^2 \frac{8}{3}\,\frac{1}{(Q r)^4}\,,\\
\label{wlrinfzero}
w_L^{(q)}(r,Q^2) &\,\xrightarrow{r \rightarrow \infty}\,
  \frac{2 N_c\alpha_\mathrm{em} Q_q^2}{\pi^2} \,Q^2 \frac{64}{15}\,\frac{1}{(Q r)^6}\,.
\end{align}
For massive quarks, $m_q> 0$, and large distances,
$r Q \gg 1$ and $r\, m_q \gg 1$, our calculation for $w_T^{(q)}$ and 
conjecture for $w_L^{(q)}$ give 
\begin{align}
\label{wtrinfmass}
w_T^{(q)}(r,Q^2) &\,\xrightarrow{r \rightarrow \infty}\,
  \frac{N_c\alpha_\mathrm{em} Q_q^2}{2\pi^2} \,Q^2  \frac{\pi m_q}{Q}\,
    \frac{ \exp(-2 m_q r)}{Q r}\,,\\
\label{wlrinfmass}
w_L^{(q)}(r,Q^2) &\,\xrightarrow{r \rightarrow \infty}\,
  \frac{2 N_c\alpha_\mathrm{em} Q_q^2}{\pi^2} \, Q^2 \frac{\pi Q}{2 m_q} \,
    \eta(m_q/Q)\, \frac{\exp(-2 m_q r)}{(Q r)^3}
\end{align}
with some $r$-independent function $\eta(m_q/Q)$ 
for which we find in the case $m_q < Q$ the numerical value
$\eta(m_q/Q)\approx 0.25$ independent of $m_q/Q$. 
The formula \eqref{wlrinfmass} is based on the assumption that 
for large $r$ the factor multiplying $\exp(-2m_q r)$ can be expanded 
in a series in $1/r$, and the leading exponent has been 
determined numerically. 

Figure~\ref{fig:dens} shows the photon densities as functions of the dipole
size $r$ together with the asymptotic expressions given above.
\FIGURE{
\includegraphics[width=0.65\textwidth]{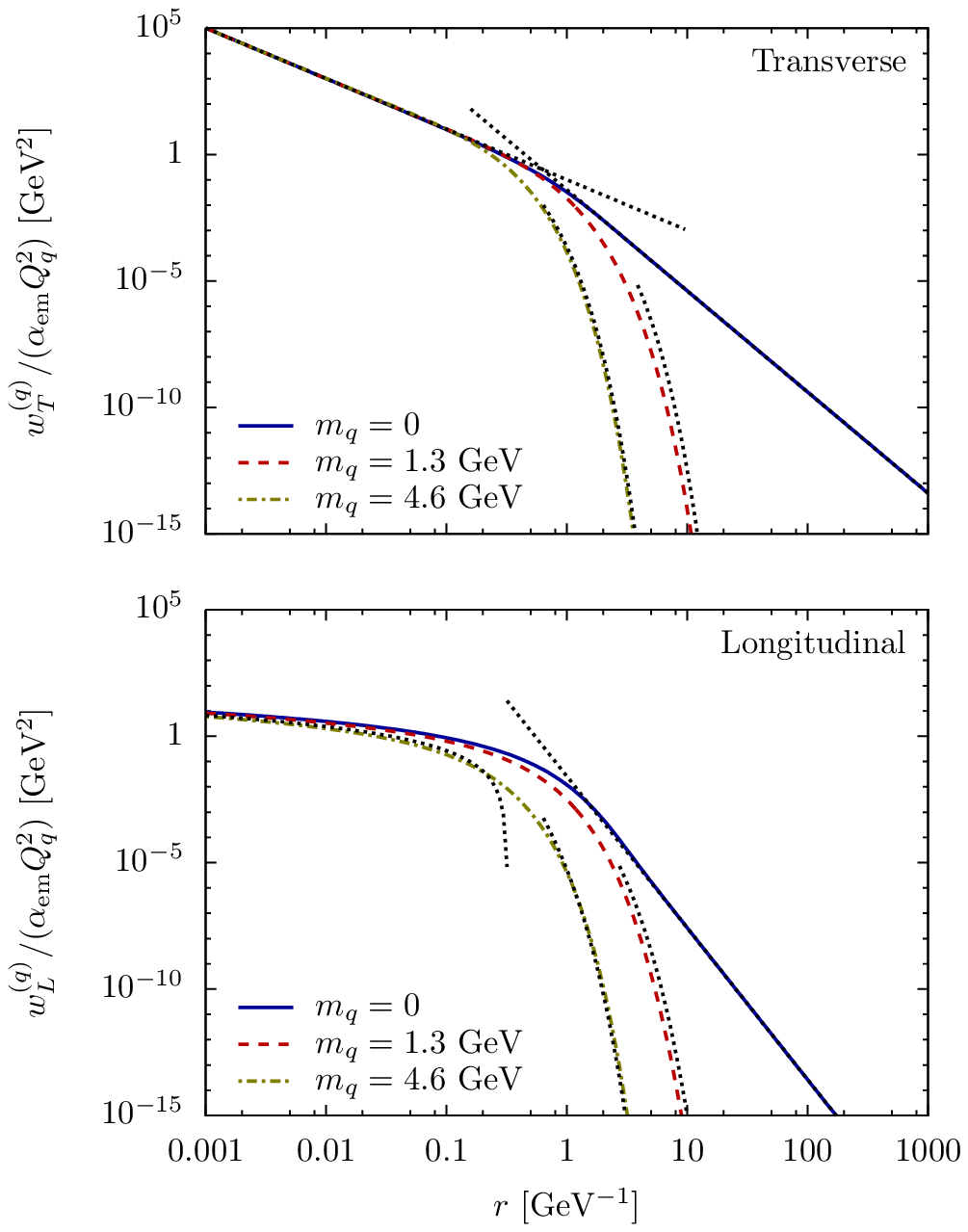}
\caption{\label{fig:dens}
Integrated photon densities $w_T^{(q)}(r,Q^2)$ (upper plot)
and $w_L^{(q)}(r,Q^2)$ (lower plot), 
both normalised to $\alpha_{\text{em}}Q_q^2$,
as a function of the dipole size $r$.
In addition, the leading terms for small and large $r$
are shown as dotted curves. 
The photon virtuality is fixed to $Q^2=10$\,GeV$^2$.
}}

\section{A simplified version of the Donnachie-Dosch model}
\label{appB}

In section \ref{sec:dipenergydep} we have used a simplified version 
of the Donnachie-Dosch model of \cite{Donnachie:2001wt}. 
In this appendix we briefly describe how the simplified model used in the 
present paper is obtained from the original model. 

We start from eq.\ (3) of \cite{Donnachie:2001wt} which is the 
cross section for the scattering of two dipoles of sizes $R_1$ and $R_2$, 
\begin{equation}
\sigma_{\text{dip}} (R_1,R_2) = 0.67 \,\frac{1}{4\pi} (\langle g^2 FF\rangle a^4 )^2 \,
R_1 \left(1- e^{-\frac{R_1}{3.1 \,a}}\right) R_2 \left(1- e^{-\frac{R_2}{3.1 \,a}}\right) 
\,,
\end{equation}
with the parameter $a=0.346 \,\mbox{fm}$ and the gluon condensate 
$\langle g^2 FF\rangle$ taken from lattice results, $\langle g^2 FF\rangle a^4= 23.77$. 
The cross section for photons or hadrons as external particles is obtained 
by folding the dipole-dipole cross section with the respective wave functions. 
This is given in \cite{Donnachie:2001wt} for the process $ab\to cb$. For our 
case of the elastic amplitude $\gamma^* p \to \gamma^* p$ these wave functions 
are those of the photon and of the proton. In \cite{Donnachie:2001wt} the latter is 
\begin{equation}
\psi_p (R) = \frac{1}{\sqrt{2 \pi} R_p}\, \exp \left(- \frac{R^2}{4 R_p^2} \right) 
\end{equation}
with $R_p=0.75\,\mbox{fm}$. 
Combining formulae (2) and (5) of \cite{Donnachie:2001wt} and relabelling $R_1$ as 
our $r$ this leads to our formulae \eqref{sigmatdip}-\eqref{densl} with 
\begin{equation}
\label{ddcross}
\sigmadip (r) = \int \ud^2 R_2 \,\abs{\psi_p(R_2)}^2 \sigma_{\text{dip}} (r,R_2) \,.
\end{equation}

In addition, an energy dependence of the dipole cross section is introduced 
by hand in \cite{Donnachie:2001wt}. 
It represents the exchanges of a soft and of a hard pomeron. The soft 
pomeron contributes only if both dipoles are larger than a certain $R_c$, chosen 
to be $R_c=0.22 \,\mbox{fm}$. The hard pomeron contributes only if at least 
one of the two dipoles is smaller than $R_c$. This leads to four different 
integration regions in the integrals over the two dipole sizes, according to whether 
$r$ and/or $R_2$ are smaller or larger than $R_c$. In our simplified model 
we assume instead that the soft pomeron is the only contribution if $r$ (the 
size of the $q\bar{q}$ pair originating from the photon) is larger than $R_c$, 
and the hard pomeron is the only contribution if $r< R_c$, independently 
of the size of $R_2$ relative to $R_c$. 

Taking into account only these two contributions with their assumed energy 
behaviour we have, similar to eq.\ (14) of \cite{Donnachie:2001wt}, 
\begin{equation}
\label{twocomp}
\sigmadip (r,W) = \theta (r -R_c) \, \sigmadip_s(r)  \left( \frac{W}{W_0} \right)^{2\epsilon_s} 
+ \theta (R_c -r)\, \sigmadip_h (r) \left( \frac{W}{W_0} \right)^{2\epsilon_h} 
\end{equation}
with the parameters given in \eqref{DDRpar}. The theta-functions indicate 
where the two contributions are relevant. 
Here $\sigmadip_s$ is given by the $R_2$-dependent factors of eq.\ (16) of 
\cite{Donnachie:2001wt}, that is by \eqref{ddcross} 
above.\footnote{Note that there is a factor $R_1 R_2$ missing in the integrals of 
equations (16) and (17) in the eprint-version of \cite{Donnachie:2001wt}. 
The journal version contains these factors. We thank H.\,G.\ Dosch 
for clarifying discussions of this point.} 
The contribution $\sigmadip_h$, on the other hand, is obtained from  
eq.\ (17) of \cite{Donnachie:2001wt} by dropping the integral over $R_1$ 
and the photon wave functions. As explained above we take into account 
only the contribution where $r<R_c$, but integrate over all dipole sizes 
in the proton. In eq.\ (17) of \cite{Donnachie:2001wt} 
this corresponds to only the second of the three integrals there, with the 
lower limit of the $R_2$-integration set to zero. Accordingly, 
\begin{equation}
\label{sigh}
\sigmadip_h (r) = \int_0^\infty 2 \pi R_2 \,\ud R_2 \,\abs{\psi_p(R_2)}^2 
\sigma_{\text{dip}} (r,R_2) \left( \frac{r}{R_c} \right)^{2 \epsilon_h} \,.
\end{equation}
The integral over $R_2$ occurring in \eqref{ddcross} and \eqref{sigh} can 
be performed numerically. Concentrating on the $R_2$-dependent factors 
only we have 
\begin{equation}
\label{oneint}
\int \ud^2 R_2 \,\abs{\psi_p(R_2)}^2 R_2 \left(1- e^{-\frac{R_2}{3.1 \,a}}\right) 
= 0.60 \,\mbox{fm} \,.
\end{equation}
Collecting all factors, we finally arrive at the simplified model \eqref{DDsig} with 
the parameters \eqref{DDRpar}. 

Finally, the original model of \cite{Donnachie:2001wt} introduces a rescaling of 
the dipole cross section by the running coupling $\alpha_s(Q^2)$, see 
eq.\ (13) there. That rescaling is relevant in particular at large $Q^2$. 
We leave out this factor in order to avoid a $Q^2$-dependence of the 
dipole cross section. 

\end{appendix}

\end{document}